\documentclass[fleqn,10pt]{wlscirep}
\usepackage[utf8]{inputenc}
\usepackage[T1]{fontenc}
\usepackage{graphicx}
\usepackage{color}
\usepackage{dcolumn}
\usepackage{latexsym}
\usepackage[normalem]{ulem}
\usepackage{hyperref,amssymb}
\usepackage{url}
\usepackage{color,soul}
\usepackage{graphicx}
\usepackage{verbatim}
\usepackage{multirow}
\usepackage{amsmath}
\newcommand{\beq}{\begin{eqnarray}}
\newcommand{\eeq}{\end{eqnarray}}
\usepackage{mathrsfs}
\usepackage{float,soul}
\usepackage{mathtools}
\usepackage{slashed}	
\usepackage{graphicx}   
\usepackage{epstopdf}
\usepackage{tabularx}
\usepackage{subfigure}  
\usepackage{hyperref}   
\usepackage{bbold}
\usepackage{wasysym}
\usepackage{feynmp}
\usepackage{hyperref}
\hypersetup{colorlinks,
}
\usepackage{dcolumn}
\usepackage{latexsym}
\usepackage{tikz}
\DeclareGraphicsExtensions{.pdf,.png,.jpg,.eps}
\newcommand{\redc}[2][red,fill=red]{\tikz[baseline=-0.5ex]\draw[#1,radius=#2] (0,0) circle ;}%
\newcommand{\bluec}[2][blue,fill=blue]{\tikz[baseline=-0.5ex]\draw[#1,radius=#2] (0,0) circle ;}%

\newcommand{\angstrom}{\textup{\AA}}
\usepackage{siunitx}
\usepackage{soul}
\usepackage{relsize}
\soulregister{\cite}7
\soulregister{\ref}7
\soulregister{\eqref}7
\usepackage{amsmath}

\usepackage{bm}
\usepackage{booktabs}

\usepackage{xcolor}
\DeclareSIUnit\angstrom{\text{\AA}}

\title{Topological defects govern plasticity and shear band formation in two-dimensional amorphous solids}

\author[1]{Xin Wang}
\author[1,2]{Jin Shang}
\author[1,3,4,*]{Yujie Wang}
\author[1,2,*]{Jie Zhang}
\author[1,5,6,*]{Matteo Baggioli}
\affil[1]{School of Physics and Astronomy, Shanghai Jiao Tong University, Shanghai 200240, China}
\affil[2]{Institute of Natural Sciences,Shanghai Jiao Tong University, Shanghai 200240, China}
\affil[3]{School of Physics, Chengdu University of Technology, Chengdu, 610059, China}
 \affil[4]{State Key Laboratory of Geohazard Prevention and Geoenvironment Protection, Chengdu University of Technology, Chengdu, 610059, China}
 \affil[5]{Wilczek Quantum Center, Shanghai Jiao Tong University, Shanghai 200240, China}
 \affil[6]{Shanghai Research Center for Quantum Sciences, Shanghai 201315, China}

\affil[*]{Corresponding authors: \color{blue}yujiewang@sjtu.edu.cn\color{black}, 
 \color{blue}jiezhang2012@sjtu.edu.cn\color{black}, \color{blue}b.matteo@sjtu.edu.cn\color{black}}

\begin{abstract}
\textbf{Understanding the fundamental mechanisms behind plastic instabilities and shear band formation in amorphous media under applied deformation remains a long-standing challenge. Leveraging on the mathematical concept of topology, we revisit this problem using two-dimensional experimental amorphous granular packings subjected to pure shear. We demonstrate that topological defects (TDs) in the displacement vector field act as carriers of plasticity, enabling the prediction of both the onset of plastic flow and the global mechanical response of the system. At the microscopic level, we show that these TDs are responsible for the dynamical breakdown of local orientational order, thereby linking topological excitations to short-range structural changes. Finally, we establish that the spatial localization and cooperative alignment of TDs underlie the formation of macroscopic shear bands, which drive plastic flow under large shear loads. This process is mediated by the quasi-periodic annihilation of large defect clusters and their geometric transformation from an approximately isotropic to a strongly anisotropic shape. Our results provide a unified framework for understanding plastic deformation and shear band formation in two-dimensional amorphous systems by connecting topology, short-range order, and mechanical response.}
\end{abstract}
\begin{document}

\flushbottom
\maketitle

\thispagestyle{empty}
\section*{Introduction}
Amorphous solids are generally characterized by the absence of long-range translational order, yet they exhibit emergent rigidity stemming from their inherent structural complexity \cite{binder2011glassy}. While structurally closer to disordered liquids, their mechanical behavior often resembles that of crystalline solids. Notably, under external loading, amorphous solids develop localized plastic activity that eventually leads to a global mechanical instability (\textit{yielding}) beyond which plastic flow emerges and the system loses its elastic response entirely \cite{Berthier2025}. Notably, the onset of plastic flow after yielding is often associated to the formation of \textit{shear bands} characterized by strong shear localization (in the context of metallic glasses, see \cite{annurev:/content/journals/10.1146/annurev.matsci.38.060407.130226}).

In contrast to crystalline materials, where plasticity and mechanical failure are well described by structural topological defects such as dislocations \cite{schmid1968plasticity,sutton2020physics}, a unified theoretical framework for the breakdown of elasticity and the emergence of irreversible plastic events in amorphous systems remains elusive. Moreover, despite significant progress (e.g., \cite{PhysRevLett.109.255502}), several questions regarding the emergence of strain localization along shear bands remain unresolved. A key challenge lies in the identification of the microscopic carriers of plasticity in glasses, which remain poorly defined due to the lack of a well-ordered reference configuration (see Fig. 1 in \cite{Baggioli2023}). In other words, the existence of well-defined defects in glasses, their underlying nature, and the possibility that they are topological in origin remain open and unresolved questions. Along the path of searching for defects in amorphous solids, two physical observables play a key role: (i) the harmonic vibrational modes and their corresponding eigenvectors (\textit{normal modes})~\cite{doi:10.1142/9781800612587_0010}, and (ii) the particle displacement field. While harmonic vibrations can, in principle, be reconstructed in real time from normal modes, the displacement field generally encodes richer dynamical information (see however a recent proposal \cite{sun2025disentangling} to define a dynamic eigenmode approach) and is more readily accessible in experiments.

In the past, the characteristics of the vibrational spectrum (normal modes) have been used to identify defects and ``soft spots'' where irreversible structural rearrangements are more frequent and microscopic plastic events take place \cite{schober1996low,widmer2006predicting,tanguy2010vibrational,manning2011vibrational}. In particular, it is now well established that amorphous solids exhibit quasilocalized excitations in their low-frequency vibrational spectrum \cite{10.1063/5.0069477}. These nonphononic modes, which coexist with phononic plane waves, are characterized by localized quadrupolar structures in the particle displacement field that can, to a good approximation, be described as Eshelby inclusions \cite{doi:10.1098/rspa.1957.0133}. Importantly, these Eshelby-like plastic events, and particularly their alignment under shear deformation, have been proposed as the driving mechanism behind shear band formation: the localization of stress along $45^{\circ}$ bands that marks the onset of plastic flow in glasses \cite{PhysRevLett.109.255502,PhysRevB.95.134111}.

\begin{figure}[ht]
    \centering
    \includegraphics[width=\linewidth]{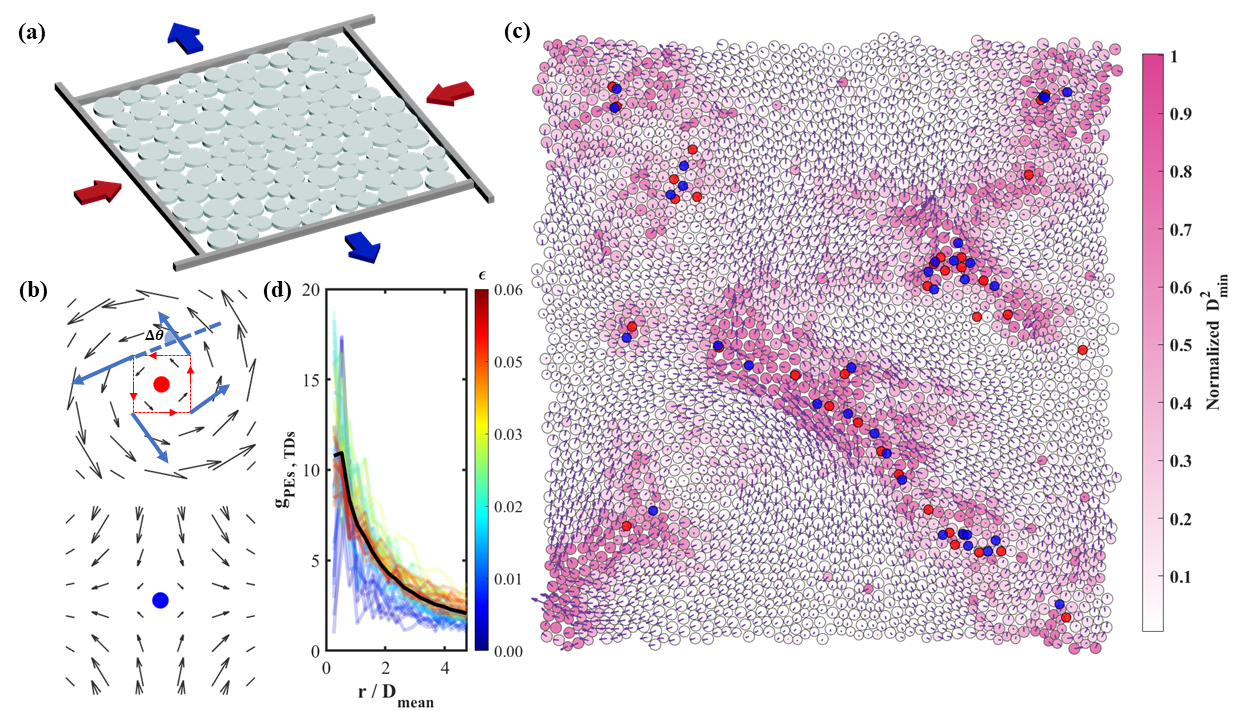}
    \caption{\textbf{(a)} Schematic illustration of the experimental setup using photoelastic disks under pure shear deformation. \textbf{(b)} Vector field corresponding to an ideal vortex (\redc{2pt}) and an ideal anti-vortex (\bluec{2pt}). In the left panel, the red box indicates the microscopic square loop used to compute the winding number $w$ using Eq. \eqref{wind2}. The blue arrows indicate the partial angle jumps $\Delta \theta_i$ on each vertex of the square loop. \textbf{(c)} The displacement vector field corresponding to the shear step $\epsilon=0.024$. Red and blue disks indicate respectively positive and negative TDs. The background color represents the local value of the non-affine squared displacement $D^2_{\text{min}}$. \textbf{(d)} The spacial correlation function between TDs and plastic events (PEs) as a function of $r/D_{\text{mean}}$. Different colors correspond to different strain values (see vertical legend) and black line is the average value over all strains.}
    \label{fig:1}
\end{figure}

The notion that vibrational modes, and their spatial structure, encode information about the defects responsible for plasticity in glasses has recently been revisited using mathematical concepts from topology. In particular, \cite{wu2023topology} demonstrated that vortex-like topological defects in the vibrational spectrum, formally defined via the topological winding number, reliably identify the plastic regions obtained using the standard non-affine metric $D^2_{\text{min}}$ \cite{PhysRevE.57.7192}. These results have been experimentally verified in a two-dimensional colloidal system \cite{Vaibhav2025} and extended also in three-dimensional simulated glasses \cite{10.1093/pnasnexus/pgae315,bera2025hedgehogtopologicaldefects3d,wu2024geometrytopologicaldefectsglasses}.

However, aside from simulations and a few exceptional experimental setups, vibrational modes, and in particular, the normal mode eigenvectors, are generally not accessible in experiments. This raises questions about the practical significance and applicability of these methods to real-world systems. In contrast, the displacement vector field, which captures particle motion relative to a reference configuration, can be directly measured in mesoscopic systems (e.g., colloids) via video microscopy, or in macroscopic systems (e.g., granular matter) using charged-coupled device cameras. It is therefore crucial to extend topological signatures beyond the abstract space of eigenvectors and into the experimentally observable displacement field. Unlike eigenvector defects, which are geometric in nature and of unclear dynamical relevance, defects in the displacement field represent \textit{bona fide} dynamical excitations.

Topological defects in the displacement field of simulated glasses were first introduced in \cite{PhysRevLett.127.015501} (see also \cite{Landry}) based on the concept of continuous Burgers vector \cite{RevModPhys.80.61}, extending the idea of dislocations to amorphous solids. This topological invariant \cite{kleinert1989gauge} has been linked to short-range structural features \cite{liu2024measurablegeometricindicatorslocal}, such as local centrosymmetry in nearest-neighbor configurations, and shown to be a powerful tool for identifying Eshelby-like plastic events in simulations \cite{bera2025burgersringstopologicalsignatures}. In parallel, vortex-like topological defects introduced in \cite{wu2023topology} have also been applied to the displacement field. Specifically, \cite{PhysRevE.109.L053002} proposed that these defects correspond to quadrupolar quasi-localized excitations (see also \cite{Hu2025} for related results), and \cite{PhysRevB.110.014107} suggested their potential role in explaining features of shear bands. We observe that in all these cases, including dislocations and disclinations in crystals, the total topological charge summed over the entire system is always zero. This highlights that crystals, and glasses more generally, are not topological phases of matter in the sense of, for example, topological insulators. What we focus on instead are local imbalances in the topological charge. These localized features may be the ones directly linked to plasticity and mechanical failure in amorphous solids. Drawing a condensed matter analogy, this scenario resembles an electronic system with a vanishing global Chern number but finite local Berry curvature \cite{10.21468/SciPostPhysLectNotes.51}.

Despite these advances, vortex-like defects in the displacement field have been experimentally studied only recently, and only in an active 2D granular system of Brownian vibrators \cite{zheng2025topologicalsignaturescollectivedynamics}, where plasticity is ill-defined due to the lack of a solid phase and external shear. Furthermore, no fundamental connection has been established between these defects, structural properties and shear band formation in glasses. We notice, however, that vortex-like rotation fields have been considered as a mechanism for shear banding in metallic glasses \cite{PhysRevLett.119.195503,SOPU2021100958}.

Here, we address this gap by analyzing a two-dimensional granular material under quasi-static shear. We show that topological defects in the displacement field are responsible for the dynamic breakdown of local orientational order, directly linking them to plastic events and soft spots. Their statistical and spatial properties align well with the global mechanical response, including the yielding transition and the emergence of plastic flow. Finally, we demonstrate that the localization and alignment of these defects, underlie the formation of shear bands. This physical process is associated with the quasi-periodic annihilation of large defect clusters and a progressive transition in their shape from isotropic to anisotropic.

We notice that a conceptually similar analysis was conducted in three-dimensional sheared granular systems \cite{Cao2018,xing2024origin}, where the defects responsible for plasticity were identified as highly distorted coplanar tetrahedra based on structural, rather than dynamical, features. Plastic events were subsequently linked to neighbor-switching processes among these defective structural units. The relation between such 3D analysis and our approach remains unclear and goes beyond the scope of this work.

\section*{Experimental setup and protocol}
The 2D granular system was composed of $2710$ small and $1355$ large photoelastic disks, with diameters $D_s = 10\ \rm mm$, $D_l = 14\ \rm mm$ and mean diameter, $D_{\text{mean}} = (D_s+D_l)/2$. 
A schematic illustration of the experimental apparatus is shown in Fig.~\ref{fig:1}(a).
The disks were placed on a horizontal glass plate, confined within a rectangular domain bounded by two pairs of movable walls. Each pair of walls could move independently, enabling the application of isotropic compression or area-conserved pure shear. 
Experimental data were acquired using a $2 \times 2$ array of high-resolution cameras (10 pixels/mm) positioned at the top of the system. 
By placing a pair of polarizers, respectively, beneath the glass plate and in front of each camera, contact forces between the disks were measured using photoelastic techniques.
For further information on the experimental setup, we refer to Refs.~\cite{wang2022experimental,Wang2021PRR}.

To prepare an initial state before shear, a stress-free, random, and homogeneous configuration was first achieved at a packing fraction $\phi = 0.834$, slightly below the jamming point $\phi_J\approx 0.84$, by adjusting the particle arrangement.
Subsequently, quasistatic isotropic compression was applied by slowly moving the boundaries inward to reach the target volume fraction $\phi=0.845$. 
During this process, we used miniature vibrators attached to the base plate to eliminate base friction, thereby ensuring an isotropic and homogeneous jammed initial state.

\begin{figure}[ht]
    \centering
    \includegraphics[width=\linewidth]{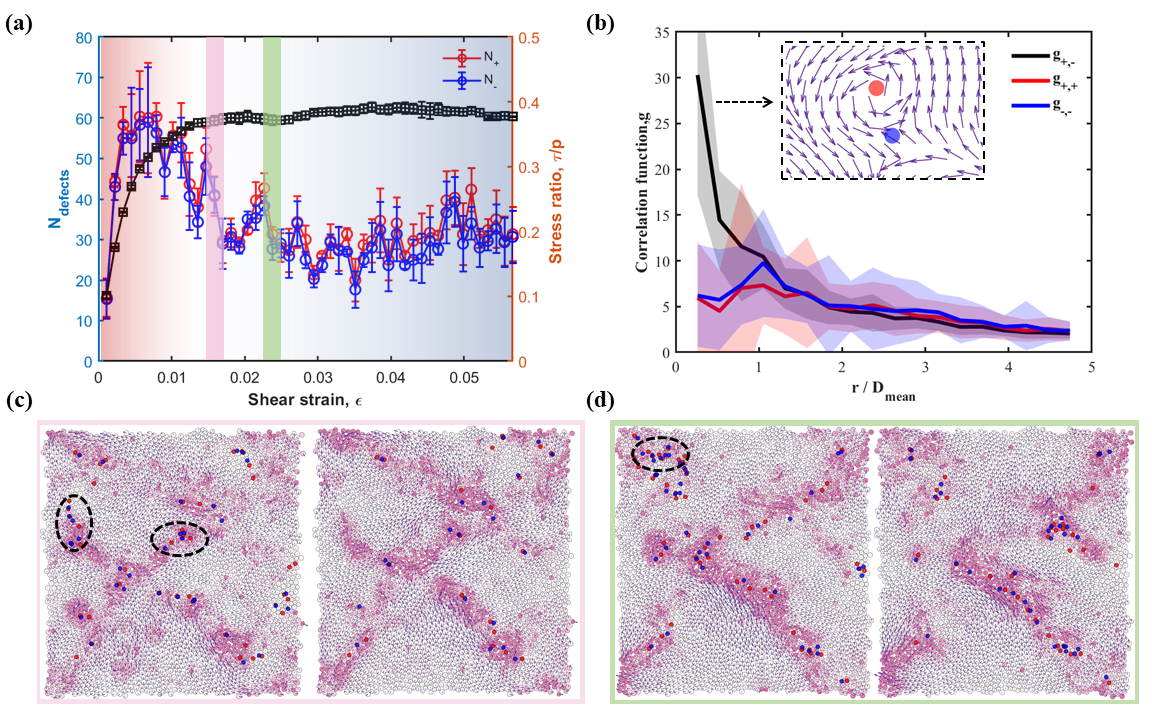}
    \caption{\textbf{(a)} The global stress ratio $\tau/p$ as a function of the shear strain $\epsilon$ (black symbols). The number of positive and negative TDs as a function of $\epsilon$ is also shown in red and blue colors respectively. Vertical color bands highlight the location of two strong drops in the number of TDs. Background colors separate the elastic regime (light red), the softening regime (white) and the steady flow regime (light blue). \textbf{(b)} The spatial correlation function between positive-positive ($++$), positive-negative ($+-$) and negative-negative ($--$) TDs as a function of the radial distance normalized by the mean particle diameter $D_{\text{mean}}$. The curves represent the average over all strain steps, while the color background bands are the corresponding error bars. The inset shows a typical vortex-pair configuration. \textbf{(c)-(d)} Comparison between the configurations before (left) and after (right) the two pronounced defect drops indicated by the vertical color bands in panel (a). Background color is the local value of $D^2_{\text{min}}$, while red and blue disks locate positive and negative vortex-like defects respectively. Black dashed loops highlight large defect clusters that disappear during the pronounced defect drops.}
    \label{fig:2}
\end{figure}

By compressing along the $x$ direction while simultaneously expanding along the $y$ direction, we gradually applied an area-conserved quasistatic pure shear deformation to the initial state. 
The shear strain defined as $\epsilon \equiv |\mathrm{d}x|/x_0$ is determined by the initial size of the system along the compression direction $x_0$ and the change in the linear size $|\mathrm{d}x|$. The maximum shear strain achieved was $\epsilon_{\mathrm{max}}=0.566$, and the entire shear process was carried out in $50$ equally spaced steps. After each step, the particle configurations and contact force networks were recorded. The pure shear protocols are identical to those employed in Ref.~\cite{shang2024yielding}.

\section*{Topological defects in the non-affine displacement field and local plasticity}

In order to define topological defects in the non-affine dynamical displacement vector field $\vec{u}=(u_x,u_y)$ (see \textit{Methods} for the precise definition), we resort to the concept of winding number \cite{kleinert1989gauge}, $w=\frac{1}{2\pi}\oint_{\mathcal{C}} d\theta$, where $\theta$ is the local orientation of the displacement field $\tan(\theta)=u_y/u_x$ and $\mathcal{C}$ is a closed loop. Whenever $\mathcal{C}$ is taken as the smallest available loop, a winding number of $w = \pm 1$ indicates the presence of a microscopic vortex (anti-vortex) excitation in the displacement field $\vec{u}$. The vector fields corresponding to a vortex defect and an anti-vortex defect are shown in Fig.~\ref{fig:1}(b). 

From a practical point of view, we project the nonaffine displacement field $\vec{u}$ on a two-dimensional square lattice grid with microscopic size $a=0.1*D_{\text{mean}}$. We then compute the winding $w$ by summing the discrete angle jumps $\Delta \theta_i$ on each side of each $a \times a$ square as show in the left panel of Fig.~\ref{fig:1}(b),
\begin{equation}
    w=\frac{1}{2\pi} \sum_{i=1}^{4}\Delta \theta_i.\label{wind2}
\end{equation}
This procedure is outlined in Fig.~\ref{fig:1}(b) for the case of an ideal vortex defect. The robustness of the method with respect to variations in the lattice grid size $a$ and the time interval used to compute the displacement field is demonstrated in the Supplementary Material (SM).

In order to elucidate the physical significance of these TDs, in Fig.~\ref{fig:1}(c), we consider a representative case corresponding to the displacement field at a strain step $\epsilon=0.024$. The vector field represents the two-dimensional displacement $\vec{u}$. The positive (vortex) and negative (anti-vortex) TDs are represented respectively with red/blue disks (\redc{2pt}, \bluec{2pt}), as illustrated in Fig.~\ref{fig:1}(b). The background color indicates the local value of the $D^2_{\text{min}}$ \cite{PhysRevE.57.7192}, calculated using the particle configurations immediately before and after the strain step (see \textit{Methods} for details). Particles with high $D^2_{\text{min}}$ are colored in purple and corresponds to plastic soft spots, areas with strong structural rearrangements.

A snapshot of the experimental system at $\epsilon = 0.024$ reveals a clear spatial correlation between topological defects (TDs) and local plastic activity, quantified using the $D^2_{\text{min}}$ metric. To further characterize this correlation, we compute the pair correlation function between TDs and plastic events (PEs), $g_{\text{PEs,TDs}}(r)$, shown in panel (d) of Fig. \ref{fig:1}. The function is evaluated by applying a threshold to identify regions with elevated $D^2_{\text{min}}$ values (see \textit{Methods} and Supplementary Material). As shown by the color legend, the correlation strength increases with strain $\epsilon$ and peaks around $\epsilon \approx 0.03$, which, as we will show, marks the onset of plastic flow. The average pair correlation function confirms a short-range correlations between TDs and PEs. Further examples and analyses of this correlation are provided in the Supplementary Material (SM). This correlation was observed in a simulated 3D amorphous system using a 2D slicing method in \cite{10.1093/pnasnexus/pgae315} and it is now confirmed by experimental inspection in a \textit{bona fide} two-dimensional system. 

We notice that, despite the TDs correspond to regions with large $D^2_{\text{min}}$, the cores of the TDs are characterized by very small values of the nonaffine displacement field $\vec{u}$. This is demonstrated explicitly in the SM. In fact, in the ideal limit, the amplitude of $\vec{u}$ has to vanish at the center of vortices and anti-vortices.
\begin{figure}[ht]
    \centering
    \includegraphics[width=\linewidth]{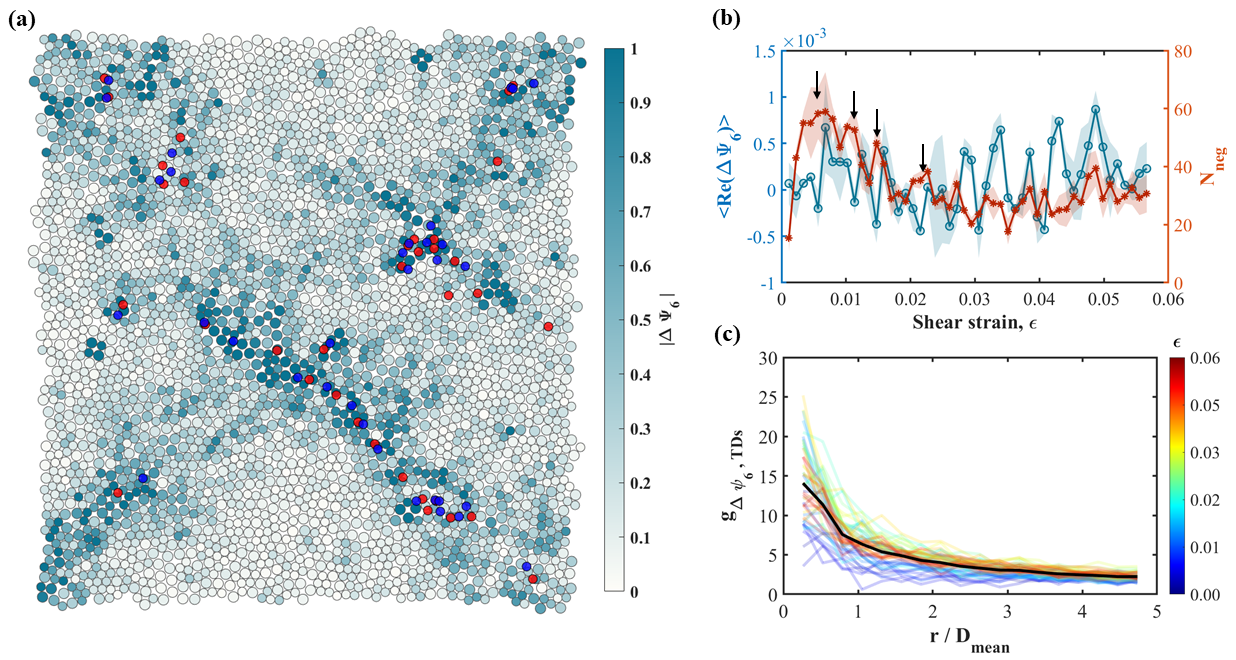}
    \caption{\textbf{(a)} A snapshot of the experimental system at $\epsilon=0.024$. Blue (\bluec{2pt}) and red (\redc{2pt}) disks represent respectively negative and positive TDs. The color map indicates the local value of the change in the sixfold orientational order $\Delta \Psi_6$. \textbf{(b)} $\Delta \Psi_6$ as a function of strain $\epsilon$ compared to the number of negative defects $N_{\text{neg}}$. Black arrows indicate evident anti-correlation at strain values corresponding to peaks in the defect count. \textbf{(c)} The spatial correlation between the regions with highest $\Delta \Psi_6$ and the TDs as a function of $r/D_{\text{mean}}$. Different colors correspond to different strain values (see vertical legend) and the black line is the average value over all strain steps.}
    \label{fig:3}
\end{figure}

These findings suggest that the TDs act as the microscopic carriers of plasticity under shear deformation, inducing local plastic flow. Using an analogous method, it was shown in \cite{zheng2025topologicalsignaturescollectivedynamics} that these defects are key elements to understand structural rearrangements and the onset of collective motion in active granular system (without external shear deformation) as well.

We notice that a correlation between vortex-like topological defects in the eigenvector field and local plasticity was initially demonstrated in \cite{wu2023topology} in a simulated 2D glass, and later experimentally confirmed in \cite{Vaibhav2025}. Nevertheless, several remarks have to be made. First, the correlation in such a case emerge only after averaging on a large number of low-frequency eigenvectors to remove the spurious effects of acoustic plane waves. Moreover, as already mentioned, and contrary to the TDs studied in our work, the topological defects in the eigenvectors are not topological excitations in glasses, but rather geometrical features of their vibrational spectrum.

We also note that, although most defects appear in pairs of opposite sign (see Fig. \ref{fig:2}(b)), some monopoles and larger structures (e.g., quadrupoles) are also observed. So far, we have not been able to determine the physical significance, if any, of these different configurations. It would be interesting to investigate this further in the future, particularly in relation to recent developments on geometric charges and anomalous elasticity in granular materials (see \cite{Kumar_2024} for a review, \cite{PhysRevE.104.024904,PhysRevE.107.055004} for more technical details and \cite{MONDAL2022112609} for experimental works in granular materials).

\section*{Topological characterization of global mechanical response}
After analyzing the dynamical displacement field and the local plastic events associated to it, we move to a global view of the mechanical properties of our experimental 2D glass.

In Fig.~\ref{fig:2}(a), we report the measured global stress ratio $\tau/p$ as a function of the shear strain $\epsilon$. Here, $\tau$ is the shear component of the stress tensor and $p$ is the pressure (see \textit{Methods} for more details). In the limit of small strain, $\epsilon \lessapprox 0.005$ (light red background), the response of the system is elastic and the stress is linear in the strain, as dictated by linear elasticity theory \cite{chaikin2000principles}. By increasing further $\epsilon$, the slope $\tau/(p \epsilon)$ monotonically decreases (softening region depicted with background white color) until the yielding point $\epsilon_Y=0.032$ where the elastic response is totally lost and plastic flow emerges (light blue background).

In the same Fig.~\ref{fig:2}(a) we plot with red and blue symbols the number of positive and negative TDs ($N_+,N_-$), observed in the displacement field at each strain step. We observe that, in the elastic regime, the number of defects grows as well with the shear strain $\epsilon$, until reaching a maximum in the region where the system starts softening and the shear modulus getting smaller. On the other hand, in the large strain region, characterized by a liquid-like plastic flow, the number of defects is approximately constant and does not vary anymore with the applied strain. 

It is interesting to note that the peak position of the topological defects ($N_+, N_-$) in Fig.~\ref{fig:2}(a) closely aligns with the peak in $\Delta Z$, the difference between the actual contact number and the critical contact number required for marginal stability, shown in Fig.~1(e) of Ref.~\cite{doi:10.1073/pnas.2402843121}.

We expect that defect creation is driven by the energy injected via shear strain and mediated by elastic forces. This energy input must overcome the nucleation cost associated with the local distortion caused by each defect. This explains why, in the elastic regime (light red background), the number of defects increases with strain.

As more defects form, additional effects begin to dominate. First, as detailed later, defects interact, introducing additional energy barriers. Second, at higher densities, the likelihood of annihilation with nearby defects increases. Third, continued deformation softens the material, reducing its rigidity and making defects more unstable and prone to disappearance. As strain $\epsilon$ increases, these effects compete with the energy injection. In the softening regime (white background), where the elastic modulus drops significantly, they dominate, leading to a decrease in defect number. We will show that this decline is also tied to the tendency of defects to localize and align along soft regions, eventually forming system-spanning shear bands.

Finally, in the plastic flow regime (light blue background), where the effective elastic modulus vanishes, defect formation ceases altogether, consistent with the complete loss of rigidity.

It is important to notice that, in very good approximation, $N_+=N_-$ at any strain step. Upon assigning a $w=+1$ charge to positive defects and a $w=-1$ charge to negative defects, this observation implies that the total winding number in the system is zero, and it remains such. This is not surprising, since this number is a topological invariant that cannot be changed by a continuous transformation, such a shear strain. This also implies that the TDs appear always in pair. 

We confirm this expectation by studying the spatial pair correlation between positive and negative defects in Fig. \ref{fig:2}(b). While defects with same charge do not show any clear structural correlations, defects with opposite charge display an evident short range correlation. 

Finally, in Fig. \ref{fig:2}(a), within the intermediate strain softening regime, we observe pronounced oscillations in the number of topological defects (TDs). To examine this behavior more closely, we focus on two significant defect drops, highlighted by the pink and green vertical bands. In Fig. \ref{fig:2}(c)-(d), we show the configurations before (left) and after (right) each drop, displaying both the vortex-like defect clusters and the local values of $D^2_{\text{min}}$.

As marked by black dashed circles, these drops are caused by the annihilation and disappearance of large defect clusters, corresponding to a sudden release of local shear stress. These dynamics are reminiscent of those reported in simulations of glasses (see \cite{10.1093/pnasnexus/pgae315}), where sharp plastic stress drops were linked to the annihilation of large clusters carrying net negative topological charge. We note that, in our case, these clusters do not exhibit a net negative charge. On the contrary, they appear to be charge-neutral on average. Finally, we observe that the periodicity of the oscillations in the defect number is intriguingly close to the peak position at $\epsilon \approx 0.005$. Moreover, although these oscillations are clearly visible in the defect number, no evident oscillatory behavior is observed in the global stress ratio. These two observations merit further investigation.

Finally, we observe that the sudden increases in the number of defects shown in Fig. \ref{fig:2}(a) are accompanied by the rapid formation of large defect clusters. These clusters coincide with the emergence of extended regions exhibiting elevated $D^2_{\text{min}}$ values (see Supplementary Material for details). Altogether, these observations suggest that the creation and annihilation of TDs is a highly collective and cooperative process, potentially linked to the phenomenon of \textit{shear avalanches} observed in granular materials and predicted by elasto-plastic mean-field models \cite{dahmen2011simple,RevModPhys.90.045006,PhysRevE.74.016118}.

\section*{Topological defects trigger dynamical breaking of local orientational order}
Having established that dynamical topological defects (TDs) in the displacement field govern both local and global plastic behavior in our two-dimensional experimental glass, a key question remains: do these defects have any relation to the underlying structure? Since amorphous solids lack long-range order, the relevant structural features must be sought in short-range order and local symmetries.

In order to do so, we resort to the concept of sixfold bond orientational order, that is described via the bond orientational order parameter:
\begin{equation}
    \Psi_6(\vec{r}_l)=\frac{1}{N_l}\sum_{j=1}^{N_l} e^{i 6 \theta_{jl}(\vec{r}_l)}.\label{bond}
\end{equation}
Here, $N_l$ is the number of neighbors of the particle located at $\vec{r}_l$ and $\theta_{lj}(\vec{r}_l)$ is the angle between the particle $l$ and the neighbor $j$ defined with respect to a fixed reference axes. More details can be found in \textit{Methods}. This complex order parameter characterizes the degree of local sixfold bond orientational order and it is equal to $1$ in a perfect triangular two-dimensional lattice.

By inspecting the static degree of short-range bond orientational order, we have not found any clear feature upon changing the strain (see SM). This suggests the absence of a direct link between local structure, at least defined by bond orientational order, and plasticity in our 2D amorphous system. To proceed, we then define the change in the bond orientational order $\Delta \Psi_6$ that corresponds to the difference between this order parameter computed before and after a strain step $\delta \epsilon$. Here, we focus on the difference in the modulus of $\Psi_6$ (see \textit{Methods}). In the Supplementary Material (SM), we also examine the difference in its phase and the modulus of its difference. Although these quantities exhibit similar features, we find that $\Delta \Psi_6$ is the most sensitive to plasticity and aligns best with the topological defects (TDs).

\begin{figure}[ht]
    \centering
    \includegraphics[width=\linewidth]{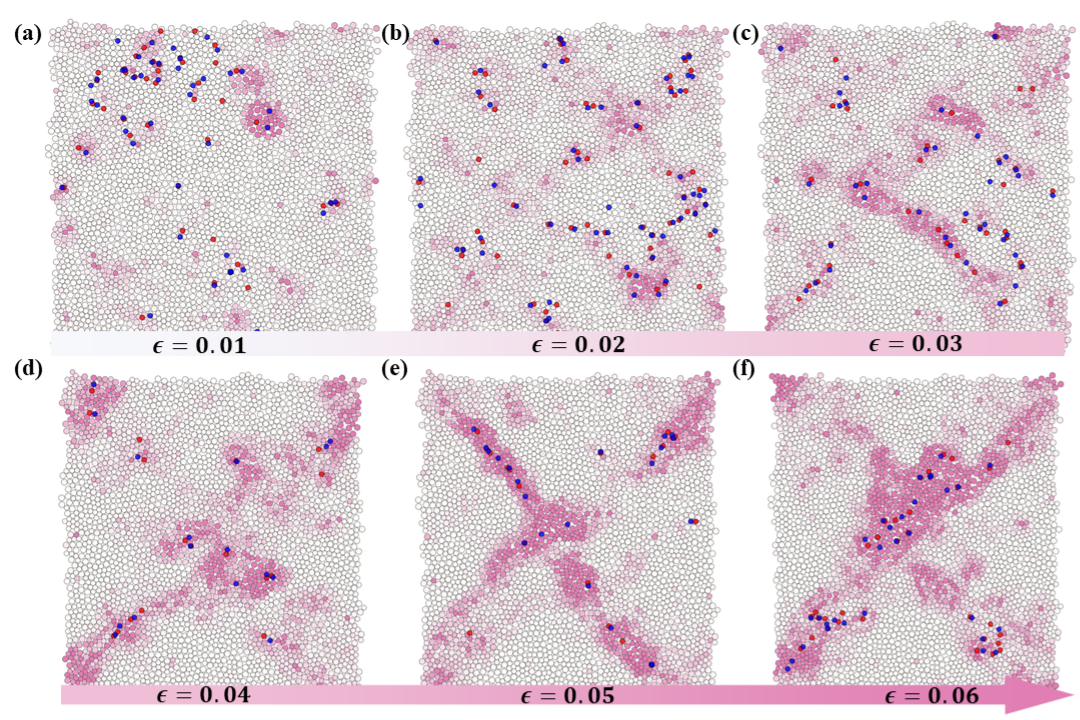}
    \caption{The evolution of the TDs as a function of the strain $\epsilon$ and the formation of shear bands. Blue (\bluec{2pt}) and red (\redc{2pt}) disks represent respectively negative and positive TDs. The color map is the local value of $D^2_{\text{min}}$ that indicates the formation of shear bands at $45^{\circ}$ degrees.}
    \label{fig:4}
\end{figure}

In Fig.~\ref{fig:3}(a), we show a snapshot of our experimental system at a strain step $\epsilon=0.024$. The background color map indicates the local value of $\Delta \Psi_6$ for each particle. Darker color represents a larger $\Delta \Psi_6$. In the same panel, we also present the location of the negative and positive topological defects. As evident, there is a clear correlation between the regions in which the change in the bond orientational order is strongest and the position of the defects. This suggests that the TDs are the microscopic responsible for the breakdown of short-range bond orientational order in our experimental system. Our results implies also that the plastic spots, characterized by a large value of $D^2_{\text{min}}$, can be also identified as the regions in which the change of the local bond orientational order is the largest, as measured by the change in the modulus of the complex order parameter $\Psi_6$.

The microscopic mechanism behind the breaking of this short-range orientational order is explained in more detail in the SM by using an ideal vortex, antivortex and vortex-antivortex pair. In most cases, a vortex mainly changes the phase angle of $\Psi_6$ while the anti-vortex its modulus. As a result, a defect pair strongly affect the bond orientational order parameter both in its direction and amplitude.

To further connect topological defects (TDs) with bond orientational order, in Fig \ref{fig:3}(b) we plot both the real part of the average $\Delta \Psi_6$ and the number of negative TDs as functions of strain. In the intermediate softening regime, a clear anti-correlation emerges: sharp drops in $\Delta \Psi_6$ coincide with peaks in the number of negative defects. This indicates that the annihilation of large TD clusters leads to pronounced changes in the average local orientational order of the 2D glass, further supporting the interpretation of TDs as key disruptors of short-range order.

Finally, values of $\Delta \Psi_6$ above a fixed threshold are selected to calculate the spatial correlation with TDs (see \textit{Methods}). As show in Fig. \ref{fig:3}(c), there is a manifest correlation between TDs and $\Delta \Psi_6$ for all the shear strain. As the shear amplitude increases, the correlation becomes stronger and then stabilizes to a stationary function after yielding. This suggests that TDs contribute to the plasticity of the system by disrupting short-range orientational order.

\section*{Shear band formation induced by localization and alignment of topological defects}
As directly demonstrated in Fig. \ref{fig:4}, upon shearing the material, and above the yielding threshold, the system develops a pair of shear bands. These bands appear as localized regions with intense shear strain and can be visualized using the $D^2_{\text{min}}$ metric. The shear bands are particularly evident in the plastic flow regime, that in our system emerges above $\epsilon \approx 0.03$, and are marked by pink color in Fig. \ref{fig:4}. Importantly, the shear bands align along $45^{\circ}$ directions.

In the past, the formation and emergent orientation of these shear bands have been ascribed to the dynamics of quadrupolar Eshelby-like structures \cite{PhysRevLett.109.255502,PhysRevB.95.134111}. Here, we revisit this question under the light of topology. We notice that vortex-like structures have been already discussed in metallic glasses in the context of shear zones and shear bands formation \cite{PhysRevLett.119.195503,SOPU2021100958} and also used phenomenologically to explain the local density variations inside shear bands \cite{PhysRevB.110.014107}. Nevertheless, a direct link to topology and the kinematics of TDs was not demonstrated.

We start by considering several snapshots of our system, that are presented in Fig. \ref{fig:4}. At low values of the strain, e.g., $\epsilon\approx0.01,0.02$ in panels (a)-(b), the structure of the TDs is rather disordered with several multi-defect clusters. By increasing further the strain $\epsilon$, shear bands start to form and TDs tend to aggregate and localize along them, i.e., in the regions where $D^2_{\text{min}}$ is largest. When the shear bands are finally formed, as seen in panels (e)-(d) of Fig. \ref{fig:4}, the TDs are strongly localized along the shear bands with almost no defect appearing elsewhere.

\begin{figure}[ht]
    \centering
    \includegraphics[width= \linewidth]{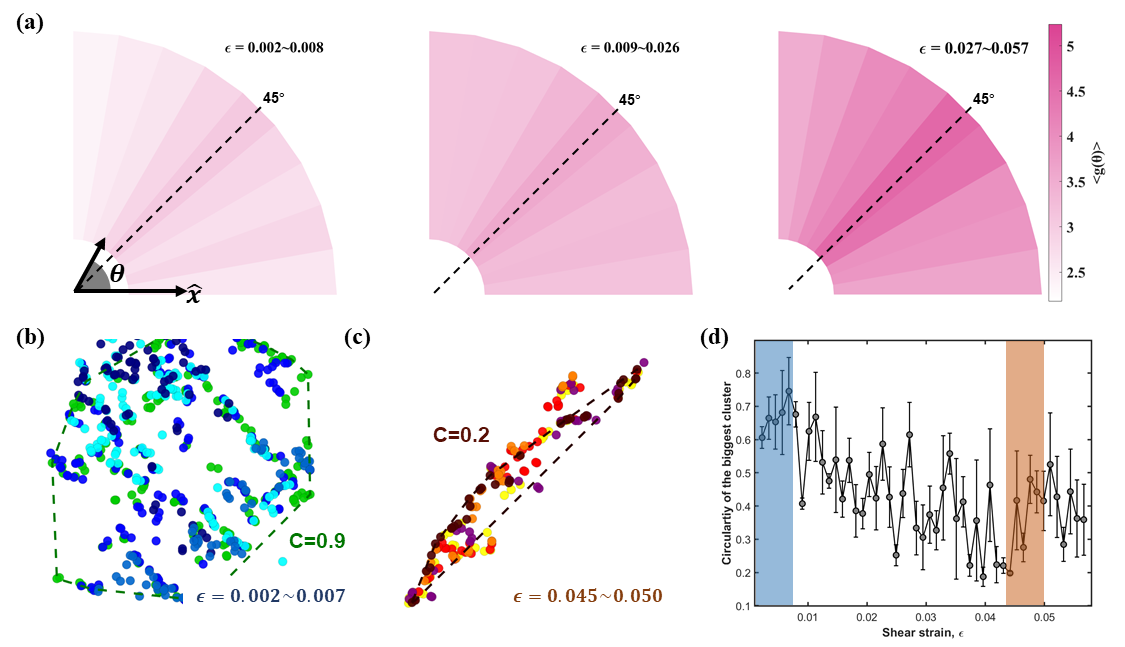}
    \caption{\textbf{(a)} The radially averaged angular pair correlation of the TDs, $g(\theta)$, in the elastic, softening and plastic flow regimes (from left to right). \textbf{(b)} The largest defect clusters in the elastic regime (blue region in panel (d)). The dashed green polygon represent the convex hull of one of the clusters corresponding to a circularity parameter $C\approx 0.9$. \textbf{(b)} The largest defect clusters in the plastic-flow regime (orange region in panel (d)). The dashed black polygon represent the convex hull of one of the clusters corresponding to a circularity parameter $C\approx 0.2$. \textbf{(d)} The circularity of the largest defect cluster $C$ as a function of the strain $\epsilon$.}
    \label{fig:5}
\end{figure}

Due to the anisotropic nature of shear, angle-averaged spatial correlations of TDs do not reveal any significant features, washing out the underlying physics. In contrast, the radially-averaged angular pair correlation function $g(\theta)$ exhibits meaningful physical behavior. We present $g(\theta)$ averaged over the shear strain $\epsilon$ in three distinct regimes corresponding to the background colors in Fig. \ref{fig:2}(a): the low-strain elastic regime, the intermediate strain-softening region, and the high-strain plastic flow regime. Our results for $g(\theta)$ are presented in Fig. \ref{fig:5}(a) focusing on $\theta \in [0,\pi/2]$.

In the elastic regime, no significant angular correlations are observed, confirming the absence of any preferred orientation in defect clusters (see panel (b) for visualizations of some of these defect clusters). In the softening region, angular correlations begin to emerge, although without a clearly favored direction. However, in the plastic flow regime, where system-spanning shear bands form (see Fig. \ref{fig:4}), a pronounced angular correlation appears at $\theta \approx 45^{\circ}$. This indicates that TD clusters become highly anisotropic, aligning along the shear bands, as is already visually evident in Fig. \ref{fig:5}. These findings support the conclusion that shear band formation is mediated by the alignment of defect clusters along the $45^{\circ}$ direction.

To further confirm the alignment of defect clusters along the shear bands, we examine their geometric properties in more detail. We then compute the convex hull of the largest defect cluster and define its circularity $C$ (see \textit{Methods} for details), which takes the value $C = 1$ for a fully isotropic shape (a perfect disk) and $C = 0$ for a one-dimensional line.

In Fig. \ref{fig:5}(b), we display the defect clusters in the low-strain elastic regime ($\epsilon \in [0.002, 0.007]$), shown in different colors. The clusters appear approximately isotropic and span the entire two-dimensional plane. The green dashed line highlights the convex hull of one of these clusters, with a circularity parameter $C \approx 0.9$, indicating a shape very close to the perfectly isotropic limit.

In contrast, Fig. \ref{fig:5}(c) presents the defect clusters in the large-strain plastic-flow regime. The clusters here are visibly elongated, strongly anisotropic, and aligned along the $45^\circ$ diagonal. The dashed polygon outlines the convex hull of one such cluster, with a circularity parameter $C \approx 0.2$, reflecting its highly anisotropic shape.

The evolution of the circularity parameter $C$ as a function of strain is shown in Fig. \ref{fig:5}(d). 

In the elastic regime, we observe a slight increase in $C$ from approximately $0.6$ to a peak around $0.9$, indicating that defects tend to cluster in roughly circular, isotropic regions. In the softening regime, where partial shear bands begin to emerge, the circularity gradually decreases, reaching a value of about $0.3$ in the plastic flow regime, beyond which it remains largely strain-independent. This trend confirms that defects not only align along the shear bands but also form highly anisotropic clusters, deviating significantly from the isotropic limit. The value of $C$ in the plastic flow regime can thus serve as an indicator of the width of the system-spanning shear bands, which is expected to depend on the interparticle potential and material properties. In the limit of very thin shear bands, we do expect $C$ to approximately vanish, as all topological defects would be aligned along a one-dimensional line at $45^\circ$.

Inspired by studies on the alignment interactions of topological defects in nematics \cite{Selinger:619845,C7SM01195D}, we present in the Supplementary Material a preliminary analysis of the interactions between TDs, including their alignment tendencies when possessing different orientations. Our current results indicate that these Coulomb-like interactions are largely governed by the total number of defects. However, due to the resolution limits of our experimental data, we are unable to accurately track the spatio-temporal patterns of the forces and torques acting on individual defects, crucial information for fully understanding their dynamics. In this context, we believe that future simulations could play a vital role in elucidating the kinematics of topological defects and the mechanisms underlying shear band formation, as suggested by our experimental findings.

\section*{Conclusions}
In this work, we investigated the plasticity of two-dimensional amorphous granular matter through the lens of topology and topological excitations. We showed that both the local and global plastic responses of the material can be understood in terms of the spatial distribution and statistical behavior of topological defects, defined via the winding number of the dynamical displacement field.

While we did not find a direct correlation between structure, specifically, bond-orientational order, and the presence of defects, we did uncover a strong link between the breakdown of this local order under deformation and the formation of defect pairs during shear. Furthermore, we observed that topological defects localize and align along shear bands. This localization suggests that these excitations not only tend to form pairs but also concentrate in regions of high strain and align with one another, possibly explaining the characteristic $45^\circ$ orientation of shear bands.

Our analysis indicates that topological defects may represent a fundamental component in understanding plasticity and mechanical deformation in two-dimensional amorphous solids.

However, much remains to be understood about their interactions and dynamics. These defects are expected to interact via Coulomb-like forces, similar to those observed for analogous topological objects in nematic liquid crystals \cite{Selinger:619845,C7SM01195D}. Interestingly, the interactions are also expected to depend on the orientation of the defects, possibly explaining their alignment along $45^\circ$ that is crucial to explain microscopically the formation of shear bands. To what extent this analogy holds, however, is still an open question. In the SM, we have presented a very preliminary exploration along these lines. A more detailed investigation of their energetics and kinematics is needed and could eventually enable a quantitative prediction of the defect number as a function of applied strain.

Moreover, it is still an open question whether the yielding transition can be interpreted as a topological phase transition mediated by these defects. So far, we have not observed any clear binding-unbinding transition of defect pairs under shear. 

Additionally, inspired by recent developments \cite{bera2025burgersringstopologicalsignatures}, it would also be promising to explore alternative topological measures, such as the continuous Burgers vector and its spatial density.

Finally, it is important to investigate how the topological properties depend on the microscopic characteristics of the system, particularly the nature of the interparticle interactions. In this context, we anticipate significant differences between materials that fail in a brittle manner and those that deform ductilely.

We leave these important questions for future investigation.

\section*{Acknowledgments}
We thank Arabinda Bera, Alessio Zaccone, Amelia Liu, Michael Moshe, Zhenwei Wu, Itamar Procaccia, Yunjiang Wang, Peng Tan and Yuliang Jin for useful discussions and related collaborations. We are particularly grateful to Walter Kob for many suggestions and comments on early results and a preliminary version of this draft.
\section*{Methods}\label{app1methods}
\subsubsection*{Topological Defects}
The displacement field can be directly calculated based on the particle positions and the strain. By averaging the positions and displacements of the particles near each boundary of the system within a distance of $1*D_{\text{mean}}$, the edge positions and displacements can be determined as $\{x_{\text{min}},x_{\text{max}},y_{\text{min}},y_{\text{max}}\}$ and $\{u_{x,\text{min}},u_{x,\text{max}},u_{y,\text{min}},u_{y,\text{max}}\}$. Using this information, the affine displacement for any position $\vec{r} = (x,y)$ inside the system is given by
\begin{align}
   &u_x^{\text{affine}}(x,y) =u_{x,\text{min}}+(u_{x,\text{max}}-u_{x,\text{min}})\,\tfrac{x-x_{\text{min}}}{x_{\text{max}}-x_{\text{min}}},\qquad \qquad u_y^{\text{affine}}(x,y) =u_{y,\text{min}}+(u_{y,\text{max}}-u_{y,\text{min}})\,\tfrac{y-y_{\text{min}}}{y_{\text{max}}-y_{\text{min}}}.
\end{align}
The nonaffine displacement is then calculated as $\vec{u} = \vec{u}_{\text{original}} - \vec{u}_{\text{affine}}$.

We interpolate $\vec u$ onto square lattice grid, as described in Eq.~\ref{wind2}. Here, a fixed lattice size $a=0.1*D_{\text{mean}}$ is chosen, which is smaller than the particle size to ensure accurate identification of the position of defects. We have verified (see SM) that the choice of $a$ does not affect our conclusions. For each $a \times a$ square loop, the directions the displacement vector at its vertices, $\theta \in (-\pi,\pi]$, are determined with respect to the $\hat{x}$ axis (see Fig. \ref{fig:1}(b) for an example). Following a clockwise loop, we compute $\Delta \theta = \theta_{i+1}-\theta_i$ (with $i$ indexing the vertices) and calculate the winding number according to Eq.~\ref{wind2}. Loops with $w$ equal to $\pm1$ are identified as TDs.

\subsubsection*{Correlation functions}
In the main text, we compute the spacial correlation functions between defects as well as between defects and regions with large variations in $\psi_6$. To identify the positions of large $\Delta \psi_6$, a threshold value is chosen as $\psi_{th} = \mathbb{E}[\Delta\psi_6] + \textbf{Std}[\Delta\psi_6]$. We select the particle position with $\Delta \psi_6 > \psi_{th}$ to calculate the spacial correlation. The correlation function is defined as:
\begin{align}
    g_{ab}(r) = \frac{S}{2\pi r \Delta r N_b N_a}\left[ \sum_{i=1}^{N_a} \sum_{j=1}^{N_b} \delta(r-r_{ij})  \right].
\end{align}
Here 'a' and 'b' correspond two the two quantities considered for the correlation (e.g., positive and negative defects or regions with large $\Delta \Psi_6$). $N_a$ and $N_b$ are the number of reference positions and $r_{ij}$ is the distance between the $i$th 'a' and $j$th 'b'.

To quantify angular correlations between particles of species $a$ and $b$, we define the full correlation function contain $r$ and $\theta$ as:
\begin{equation}
g_{ab}(r, \theta) = \frac{S}{r\, \Delta r\, \Delta \theta\, N_a N_b} \sum_{i=1}^{N_a} \sum_{j=1}^{N_b} \delta(r - r_{ij})\, \delta(\theta - \theta_{ij}),
\end{equation}
where $r_{ij}$ and $\theta_{ij}$ denote the relative distance and angle between particle $i \in a$ and particle $j \in b$, respectively. $\Delta r$ and $\Delta \theta$ are the corresponding bin widths in radial and angular directions, and $S$ is the area we considered . 

To isolate the angular component, we define the angle-resolved pair distribution function by summing over the radial direction within a specified range:
\begin{equation}
g_{ab}(\theta) = \sum_{r_k = r_{\min}}^{r_{\max}} g_{ab}(r_k, \theta)\, \Delta r,
\end{equation}
which represents the probability of finding a particle of species $b$ at a relative angle $\theta$ with respect to a particle of species $a$, averaged over a selected radial shell.

\subsubsection*{Local Plasticity $D^2_{\text{min}}$}
The locally irreversible rearrangements of particles are evaluated by the minimum of mean-square difference, $D_\text{min}^2$, between the actual displacements of the neighboring particles relative to the central one and the relative displacements that they would have if they were in a region of uniform strain $\epsilon_{ij}$. We define
\begin{align}
    D^2(t,\Delta t) = \sum_n \sum_i \Bigl(  r_n^i(t)-r_0^i(t) -   \sum_j (\delta_{ij}+\epsilon_{ij})\times[r_n^j(t-\Delta t)-r_0^j(t-\Delta t)]  \Bigr)
\end{align}
The uniform strain $\epsilon_{ij}$ is chosen such that it minimizes $D^2$. More information can be found in the original Ref. \cite{PhysRevE.57.7192}. 

\subsubsection*{Local Orientational Order $\Delta \Psi_6$}
To describe the local orientational order, we use the complex sixfold order parameter $\Psi_6$, as defined in Eq. \eqref{bond} in the main text. The neighbors $\vec r_j$ of an arbitrary central particle $\vec r_l$ are found by Delaunay triangulation for the center of each particle in system. Then we obtain the phase angle for each neighbor (labeled by $\vec r = \vec r_j - \vec r_l$): $\theta_{lj} = \arctan (r_y,r_x)$. Here the $\arctan(y,x)$ is a four-quadrant inverse tangent so there is $\theta_{lj} \in (-\pi,\pi]$. For each strain we can get the $\Psi_6(t)$ for all the particles.

To reveal the variation in local orientational order, we define the difference as $\Delta \Psi_6(t) = |\Psi_6(t)| - |\Psi_6(t - \Delta t)|$ where $\Delta t$ corresponds to a single strain step.

\subsubsection*{Circularity of Defect Clusters}
In order to construct the adjacency matrix, we define a connection between two particles if their distance along a prescribed direction $\theta_{s}$ is less than $15 D_{\text{mean}}$ and their perpendicular distance is less than $3 D_{\text{mean}}$. We select $45^\circ$ and $135^\circ$ as characteristic directions, then find the largest cluster to analyze. 

The shape of each cluster is quantified by the circularity of its convex hull, defined as
$C = \frac{4\pi A}{P^2}$
where $A$ and $P$ are the area and perimeter of the convex hull. For a perfect circle, \(C = 1\), while for an ideal line segment, \(C = 0\).

\subsubsection*{Stress Measurements}
At each strain value $\epsilon$, we capture both a stress image using a polarizer in front of the cameras and a normal image without the polarizer to detect the positions of the disks. The contact forces are measured with an accuracy of 5\% using a force-inverse algorithm detailed in Ref. \cite{Wang2021PRR}.

By utilizing these contact forces, the system's stress tensor $\hat{\sigma}$ can be defined as $\hat{\sigma} = \frac{1}{S} \sum_{i \ne j} \mathbf{r}_{ij} \otimes \mathbf{f}_{ij},$
where $S$ represents the system's area, $\mathbf{r}_{ij}$ denotes the contact vector from the center of particle $i$ to the contact point between particles $i$ and $j$, $\mathbf{f}_{ij}$ is the force vector associated with this contact, and $\otimes$ signifies the vector outer product. The principal stresses of $\hat{\sigma}$ are denoted as $\sigma_1$ and $\sigma_2$, with the pressure $p$ defined as $p = (\sigma_1 + \sigma_2)/2$ and the shear stress $\tau$ as $\tau = |\sigma_1 - \sigma_2|/2$.

\subsection*{Data availability}
The datasets generated and analyzed during the current study are available upon reasonable request by contacting the corresponding authors. 

\subsection*{Funding}
MB acknowledges the support of the Shanghai Municipal Science and Technology Major Project (Grant No.2019SHZDZX01) and the support of the sponsorship from the Yangyang Development Fund. JS and JZ acknowledge the support of the NSFC (No.11974238 and No.12274291) and the Shanghai Municipal Education Commission Innovation Program under No. 2021-01-07-00-02-E00138. JS and JZ also acknowledge the support from the SJTU Student Innovation Center. XW and YW are supported by the National Natural Science Foundation of China (No. 12274292).

\subsection*{Author contributions statement}
MB conceived the idea of this work. JS performed the experiments with the help of JZ. XW performed the numerical analysis and theoretical modeling with the help of MB and YW. MB and XW wrote the manuscript with the help of all other authors. All authors participated in the discussion of the results and their physical interpretation.

\appendix 
\renewcommand\thefigure{S\arabic{figure}}    
\setcounter{figure}{0} 
\renewcommand{\theequation}{S\arabic{equation}}
\setcounter{equation}{0}
\renewcommand{\thesubsection}{SM\arabic{subsection}.}
\clearpage
\newpage
\section*{\Large Supplementary Material}

\subsection{Non-affine displacement and topological defects}
\begin{figure}[h]
    \centering
    \includegraphics[width=\linewidth]{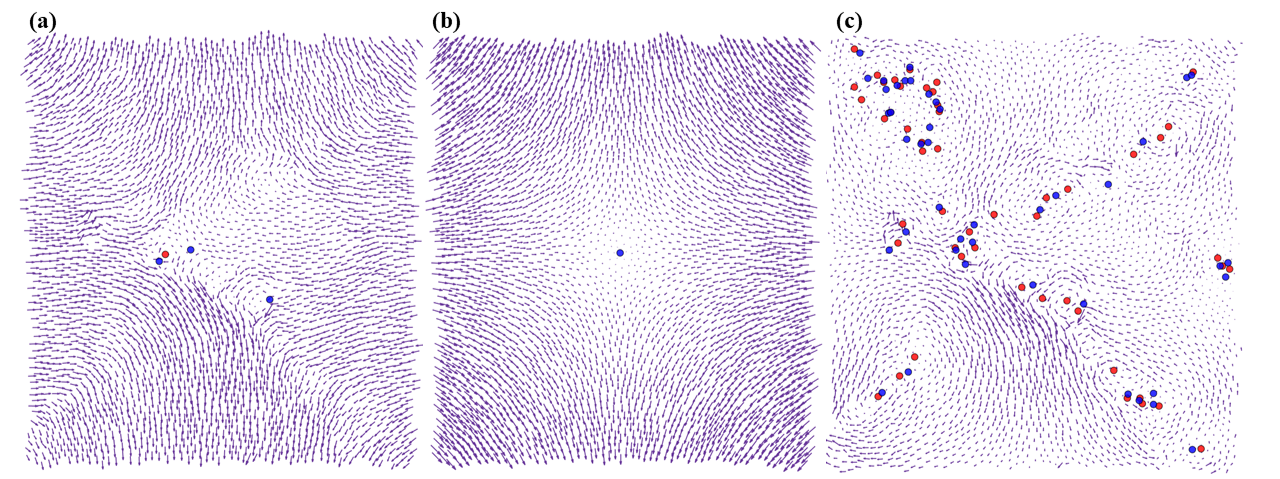}
    \caption{Topological defects in: \textbf{(a)} original displacement field, \textbf{(b)} affine displacement field, \textbf{(c)} nonaffine displacement field. All images refer to $\epsilon = 0.0226$.}
    \label{si1fig1}
\end{figure}

In our work, we define topological defects based on the non-affine displacement field, as this is the physically relevant quantity for discussing plastic deformations. In Fig. \ref{si1fig1}, we provide additional details supporting this choice, using a benchmark case corresponding to $\epsilon = 0.0226$.

Panel (a) of Fig. \ref{si1fig1} shows the original experimental displacement field, where very few topological defects are observed. In panel (b), we isolate the affine component of the displacement vector, computed using the method described in the \textit{Methods} section. Here, we observe only a single negative topological defect, which is a direct consequence of the global affine pure shear deformation.

In contrast, panel (c) shows the non-affine displacement field, obtained by subtracting the affine component from the total experimental field. In this case, numerous topological defects become clearly visible. It is important to emphasize that the calculation of the winding number is not a linear operation with respect to the addition or subtraction of vector fields. 

\begin{figure}[h]
    \centering
    \includegraphics[width=\linewidth]{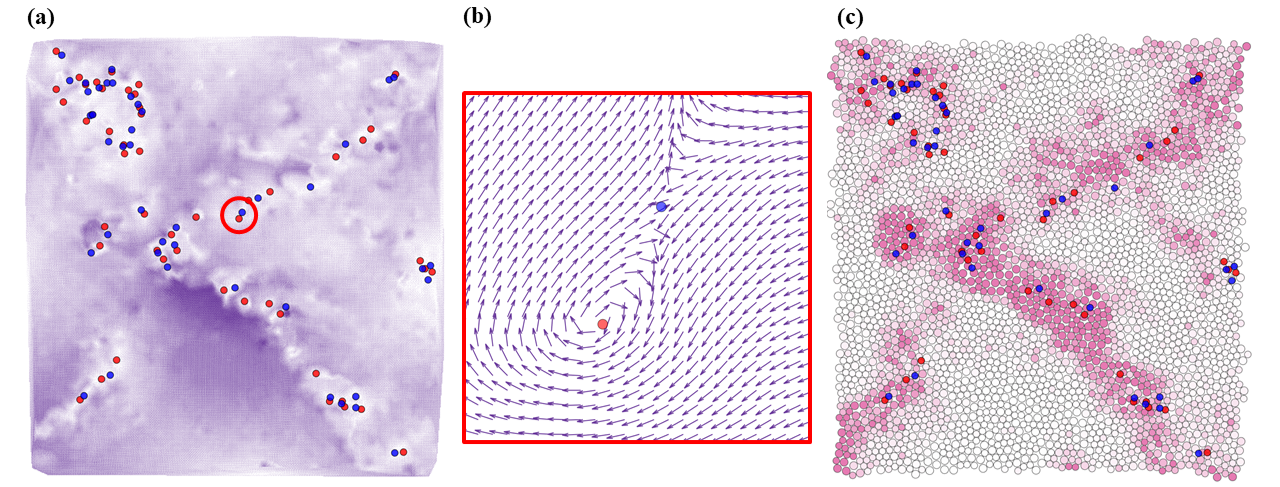}
    \caption{\textbf{(a)} Interpolated nonaffine displacement field (purple color). Red and blue disks are topological defects with $w =\pm 1$. The region within the red circle is magnified and shown in panel \textbf{(b)}. This close-up allows us to verify that the defects observed under experimental conditions are consistent with those expected in the ideal case. \textbf{(c)} Local map of $D^2_{\text{min}}$. Comparing panels (a) and (c) it is evident that the smaller amplitude of the displacement field, the larger $D^2_{\text{min}}$, and the more the defects in the corresponding region.}
    \label{si1fig2}
\end{figure}

In Fig. \ref{si1fig2}(a), we show the interpolated non-affine displacement field overlaid with the topological defects (TDs). It is evident that TDs tend to appear in regions where the displacement field is small or nearly vanishing. This observation aligns with the general expectation that, at the core of a defect, regardless of its winding number, the magnitude of the associated vector field must vanish.

For clarity, Fig. \ref{si1fig2}(b) provides a zoomed-in view of a pair of topological defects with opposite charges and their corresponding vector field. Finally, panel (c) demonstrates that regions with high values of $D^2_{\text{min}}$ tend to correlate with areas where the non-affine displacement field is small, precisely where most TDs are located.

\begin{figure}[h]
    \centering
    \includegraphics[width=0.9\linewidth]{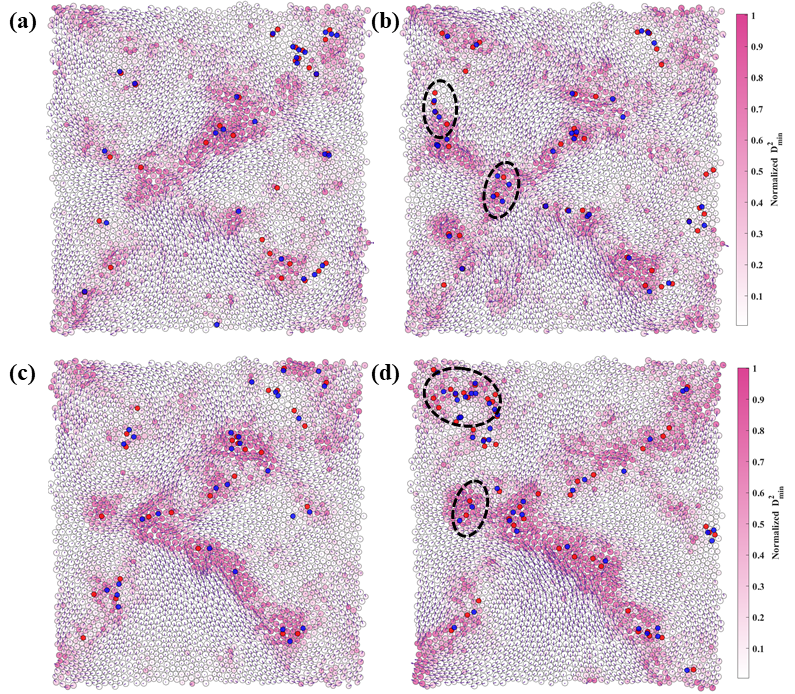}
    \caption{The creation of large defect clusters corresponding to the sudden increases in defect numbers observed in Fig. \ref{fig:2}(a) in the main text. \textbf{(a-b)} and \textbf{(c-d)} correspond to the transition between $\epsilon = 0.0134 \to 0.0136$ and $\epsilon = 0.0193 \to 0.0227$. The black dashed circles mark the emergence of large defect clusters during those strain steps.}
    \label{si1fig3}
\end{figure}

In Fig. \ref{si1fig3}, we present additional evidence for the emergence of large defect clusters at strain levels where the number of defects sharply increases. These snapshots also offer further visual confirmation of the spatial correlation between TDs and PEs.

\subsection{Correlation between plastic events and topological defects}
\begin{figure}[h]
    \centering
    \includegraphics[width=0.85\linewidth]{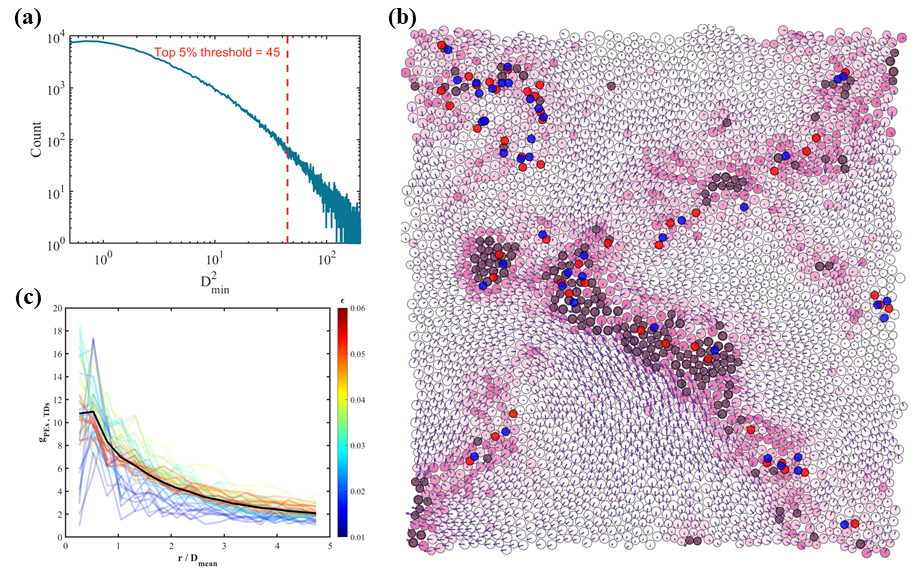}
    \caption{\textbf{(a)} Distribution of $D^2_{\text{min}}$ for all particles and all values of shear strain. The red dashed line labels the threshold of top 5\% particles with large $D^2_{\text{min}}$ used to define the plastic soft spots. \textbf{(b)} A snapshot of the non-affine field for $\epsilon = 0.0226$. The gray particles represent those with $D^2_{\text{min}}$ values exceeding the threshold, and are identified as the locations of plastic events (PEs). Blue and red disk are the TDs. \textbf{(c)} The spatial correlation between TDs and PEs. Different colors refer to different values of the shear strain $\epsilon$ (color bar legend), while the black line is the average over all $\epsilon$.}
    \label{si2fig1}
\end{figure}

To quantitatively characterize the results shown in Fig. \ref{fig:1} in main text, we define plastic events (PEs) and use them to compute their spatial correlation with topological defects, $g_{\text{PEs},\text{TDs}}$. As shown in Fig. \ref{si2fig1}(a) we collected the \( D_{\text{min}}^2 \) values of all particles over the entire range of strain steps for statistical analysis and used the resulting distribution to determine a threshold for identifying plastic events. Particles with larger \( D_{\text{min}}^2 \) result having correlation with TDs, as visually evident in Fig.\ref{si2fig1}(b). We provide further analysis of this correlation in Fig\ref{si2fig1}(c), where the radial pair correlation function between PEs and TDs is shown for all values of the shear strain $\epsilon$, including the average value (black curve). The short-range correlation is evident, and it is larger for larger values of the strain and in particular close to the onset of plastic flow.

\begin{figure}[h]
    \centering
    \includegraphics[width=\linewidth]{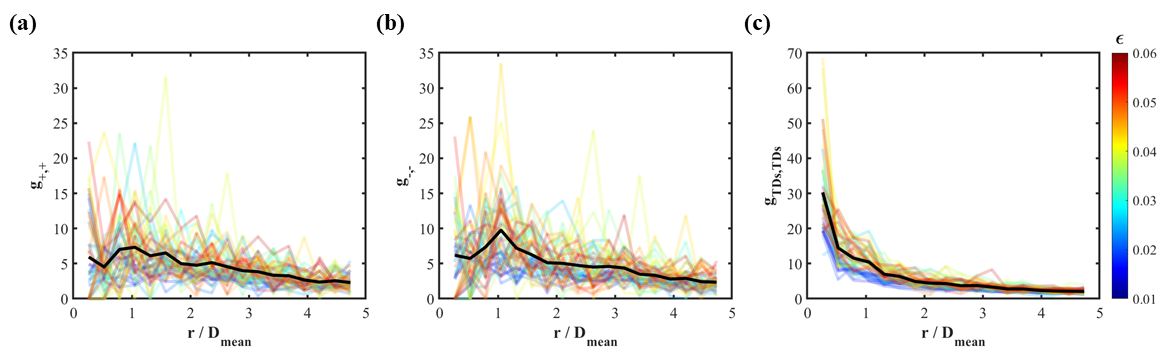}
    \caption{Radial pair correlation between topological defects (TDs). \textbf{(a)} Correlation between positive defects. \textbf{(b)} Correlation between negative defects. \textbf{(c)} Correlation between TDs, independently of their winding number. Different colors refer to different values of the shear strain $\epsilon$ (see color legend). Black curve is the average computed on all values of $\epsilon$.}
    \label{si2fig2}
\end{figure}

In Fig. \ref{fig:2}(b), we have presented the average correlation between TDs. Here, we separately analyze the correlations between different types of defects. The results are presented in Fig.\ref{si2fig2}. The spatial correlation between defects of the same type is nearly constant, with a mild peak around $r\approx D_{\text{mean}}$. In contrast, the correlation between defects without distinguishing their charge (panel (c)), or defects with opposite charge (Fig. \ref{fig:2}(b) in the main text), is significantly stronger, suggesting that defects tend to appear in pairs. In summary, our analysis suggests that defects tend to form 
'$-+-$' or '$+-+$' sequences over a particle length scale, which also supports the presence of local alignment behavior.

\subsection{Bond orientational order under shear deformation}
\begin{figure}[h]
    \centering
    \includegraphics[width=0.95\linewidth]{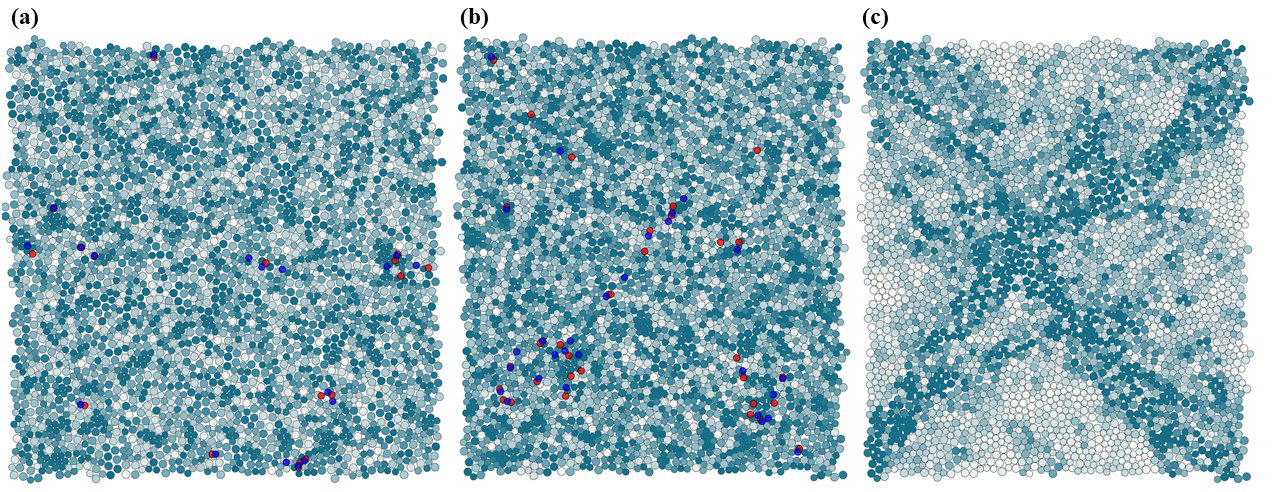}
    \caption{The spatial distribution of $\Psi_6$ at \textbf{(a)} $\epsilon=0$ and \textbf{(b)} $\epsilon=0.0566$ (the maximum value of the shear strain). Blue and red dots are TDs.  \textbf{(c)} Distribution of $\Delta\Psi_6$ for $\Delta \epsilon= 0.0566$.}
    \label{si3fig1}
\end{figure}

Based on the definition of $\Psi_6$ in the main text, Eq. \eqref{bond}, we compute its distribution in Fig.\ref{si3fig1}(a) and (b) for two different values of the strain. No correlation is detected between TDs and the local value of \( \Psi_6 \). The spatial distribution of \( \Psi_6 \) shows no discernible pattern across all strain levels. This implies that TDs in the displacement field bare no relation to the static structure of our glass.

\begin{figure}[h]
    \centering
    \includegraphics[width=0.8\linewidth]{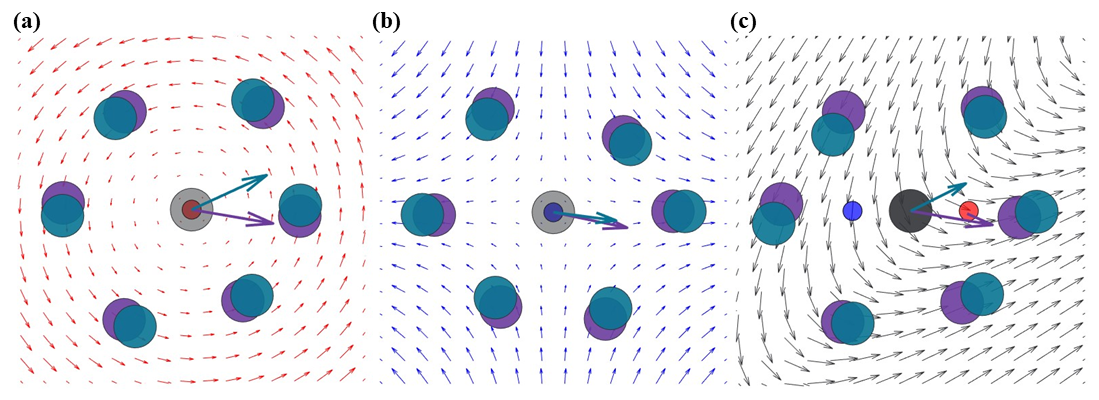}
    \caption{$\Delta\Psi_6$ induced by a positive vortex \textbf{(a)}, a negative one \textbf{(b)} and a pair of TDs with opposite sign \textbf{(c)}. The gray disk is the center particle, while purple disks its first neighbors before the appearance of a TD. The green disks represent the position of the first neighbors induced by the displacement associated to the TD. The colored arrows show the nonaffine displacement fields. The purple and green arrow represents $\Psi_6$ before the displacement and $\Psi_6$ after the displacement induced by the TD.}
    \label{si3fig2}
\end{figure}

However, when analyzing the change in the local orientational order, \( \Delta\Psi_6 \), the shear band structure becomes clearly visible, as shown in Fig. \ref{si3fig1}(c). Therefore, we use \( \Delta\Psi_6 \) as a measure of local orientation change to further explore the connection between TDs and local plasticity. In Fig.\ref{si3fig2}, we demonstrate how the displacement field of an idealized topological defect influences the local value of \( \Psi_6 \). The complex order parameter \( \Psi_6 \) is represented as a vector to simultaneously convey its magnitude and phase. We find that positive defects primarily alter the phase of \( \Psi_6 \), while negative defects mainly affect its magnitude. A typical defect pair in the system can thus induce a significant distortion in the \( \Psi_6 \) field. Building on this, we separately examine the variations in the phase and magnitude of \( \Psi_6 \).

\begin{figure}[h]
    \centering
    \includegraphics[width=\linewidth]{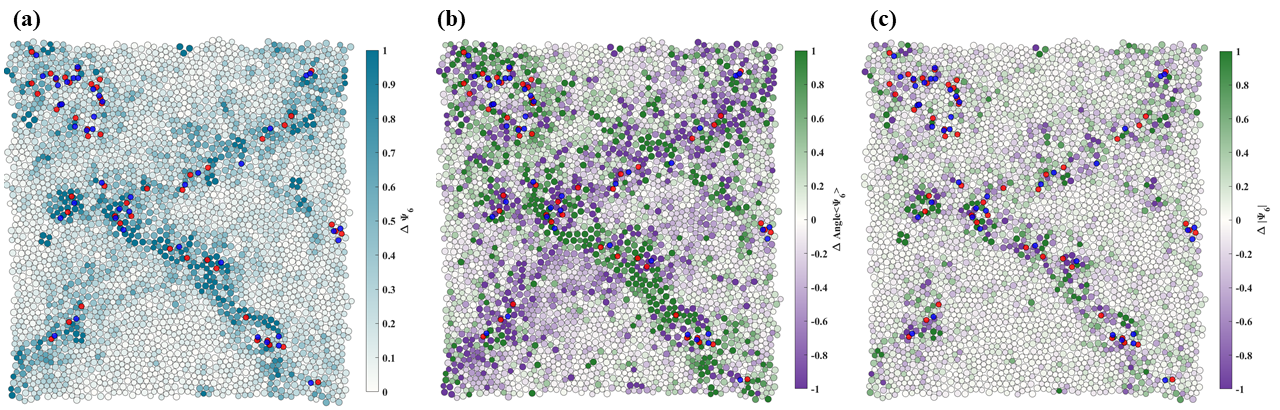}
    \caption{The distribution of \textbf{(a)} $\Delta\Psi_6$, \textbf{(b)} phase angle of $\Delta\Psi_6$ and \textbf{(c)} modulus of $\Delta\Psi_6$ at $\epsilon=0.0226$. The color intensity represents the magnitude, as outlined by the different color bar legends.}
    \label{si3fig3}
\end{figure}

Fig.\ref{si3fig3} shows the spatial distribution of the absolute value of $\Delta\Psi_6$ (panel (a), same data as Fig. \ref{fig:3}(a) in the main text), the change of the phase angle of \( \Psi_6 \) (panel (b)), and the change of the absolute value of \( \Psi_6 \) (panel (c)).  All three quantities reveal the presence of the shear band at $\epsilon=0.0226$. The variation of the phase angle in Fig\ref{si3fig3}(b) exhibits different rotational tendencies along the two shear bands in different direction, which can be associated with the shear geometry of the system. In particular, on one of the bands, its value is negative (purple color), while on the other is positive (green color).

\begin{figure}[h]
    \centering
    \includegraphics[width=\linewidth]{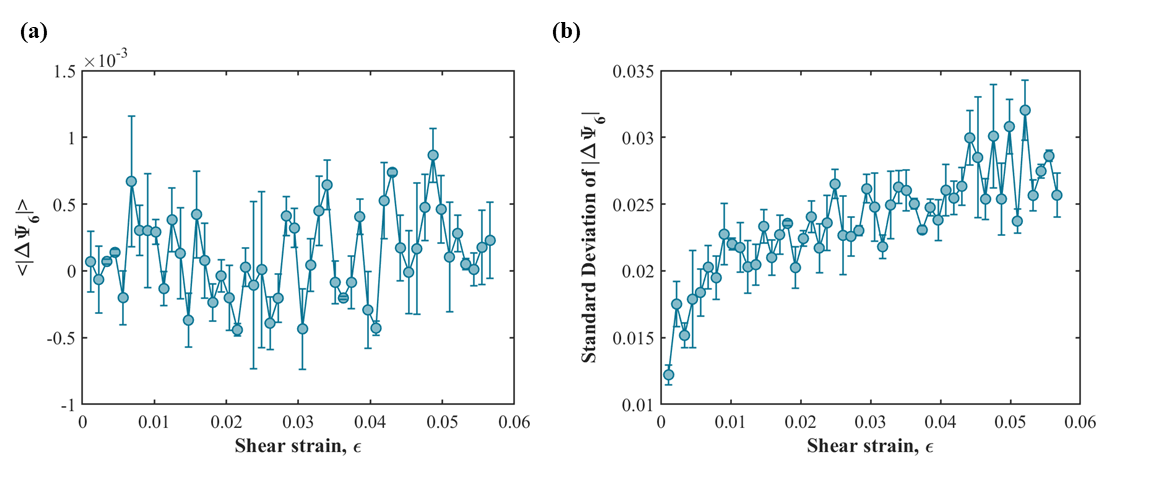}
    \caption{Evolution of $\Delta \Psi_6$ with shear strain. \textbf{(a)} The absolute value of $\Delta \Psi_6$. \textbf{(b)} The standard deviation of $\Delta \Psi_6$.}
    \label{si3fig4}
\end{figure}

Finally, we examine whether the distribution of \( \Delta\psi \) exhibits any quantitative patterns in Fig.\ref{si3fig4}. The mean value (panel (a)) shows no clear trend, but its oscillatory pattern match the one reported in the number of defects, as noted in the main text. In contrast, its standard deviation closely follows the stress-strain curve. Taken together, these results demonstrate an evident correlation between \( \Delta\Psi_6 \) and the plasticity at both macroscopic and microscopic levels. Through its connection with TDs, \( \Delta\Psi_6 \) serves as an effective bridge linking topological defects to plastic deformation.

\subsection{Cluster analysis and circularity}
For the definition of clusters, we adopt a specific method for constructing the adjacency matrix. In the conventional approach, the adjacency matrix is constructed based solely on inter-particle distance. However, this often causes shear bands in two different directions to merge at their intersection, resulting in a single connected structure. Such merging complicates the subsequent analysis of shear band formation and evolution. Therefore, we construct the adjacency matrix by connecting two particles only if they lie within the same elongated rectangular region ($3D\times15D$). Based on the symmetry of the system, clusters are selected along both the 45$^\circ$ and 135$^\circ$ directions. For further analysis, we focus on the largest cluster in each direction, defined as the one containing the highest number of defects.

\begin{figure}[h]
    \centering
    \includegraphics[width=0.9\linewidth]{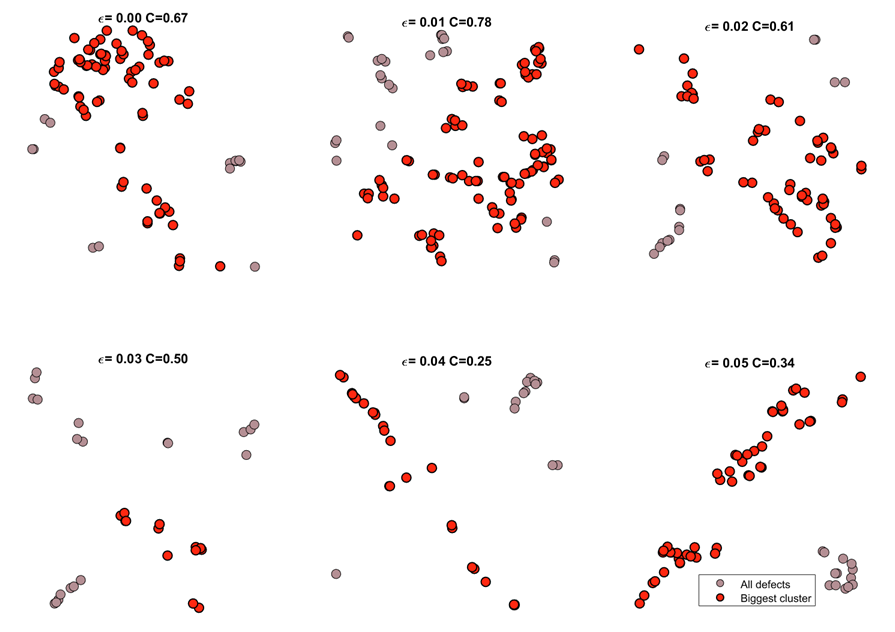}
    \caption{Identification of the largest defect clusters at different strain. The values in each panel show the shear strain $\epsilon$ and the circularity parameter of the largest cluster.}
    \label{si4fig1}
\end{figure}

This approach effectively captures the shear bands in the system as shown in Fig.\ref{si4fig1}(d-f) and also effectively distinguishes the clusters that appear prior to shear band formation as shown in Fig.\ref{si4fig1}(a-c).

\begin{figure}[h]
    \centering
    \includegraphics[width=0.7\linewidth]{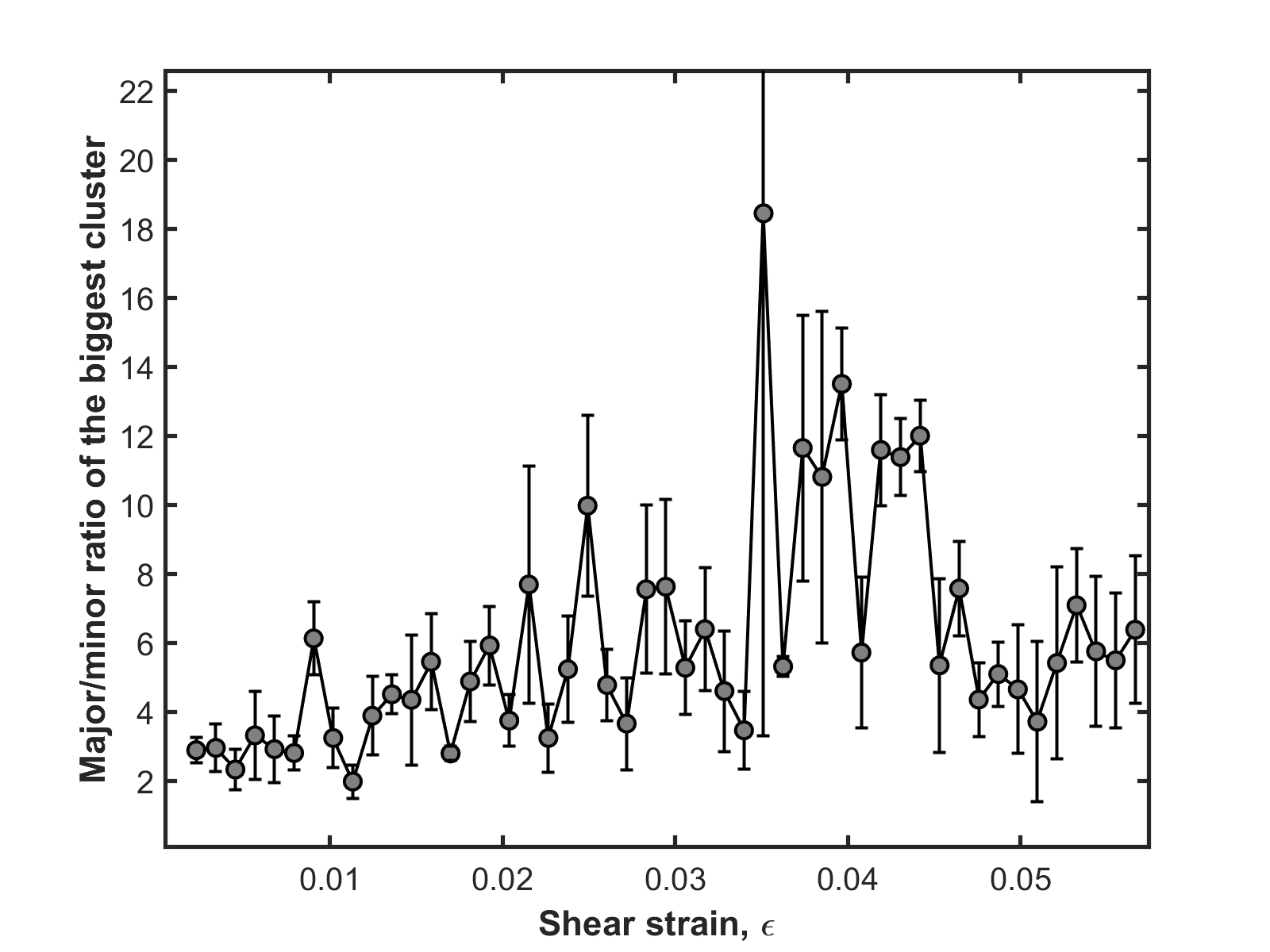}
    \caption{The ratio of the major to minor axis as a function of shear strain.}
    \label{si4fig2}
\end{figure}

To quantitatively characterize the shape of clusters, we define a circularity measure $C$ (see \textit{Methods} and main text) which is directly indicated in each panel of Fig.\ref{si4fig1}. Additionally, we calculate the major and minor axes of each cluster and use their ratio to describe the degree of elongation as shown in Fig.\ref{si4fig2}. These results suggest that the largest cluster becomes increasingly elongated following shear band formation. 

\subsection{Energy of topological defects}
In order to understand the localization and alignment of the vortex-like TDs, it is fundamental to notice that the latter are not only characterized by a winding number $w$, Eq. \eqref{wind2}, but also by an orientation \cite{Selinger:619845}. In particular, in polar coordinates $(r,\phi)$, a topological defect in two dimensions can be parametrized as
\begin{equation}
    \theta(r,\phi)= w \phi+ \theta_0 .
\end{equation}
Here, $w$ is the topological winding number defined in Eq. \eqref{wind2} while $\theta_0$ contains information about the orientation of the TD.
In the case of an anti-vortex with $w=-1$, the orientation of the defect can be represented by a unit vector $\hat{p}\equiv (\cos \varphi, \sin \varphi)$ where $\varphi=\theta_0/2\,(\text{mod}\,\pi)$. The defect structure has twofold rotational symmetry, and the vector field points outwards along the direction of $\hat{p}$ and its $\pi$-transform (see Fig. 6.1 in \cite{Selinger:619845}). Different is the case of a $w=+1$ vortex defect for which orientation is not well-defined. In fact, in that case $\theta_0$ does not really represent the orientation of the defect but rather the structure of the vector field around it (radially outward or inward, tangential counter-clockwise or clockwise, or intermediate). See \cite{C7SM01195D} for more details.

Building on the analogy with liquid crystals, superfluids and magnets, we do expect the TDs to interact via two-dimensional Coulomb like interactions. Nevertheless, differently from electric charges in two-dimensional electromagnetism, these defects are endowed with an orientation. Hence, their interaction is more complex and it is strongly affected by their relative orientation. The analytical derivation of the interaction potential and the alignment potential can be obtained using the mathematical method of conformal mapping (see \cite{C7SM01195D} for details). Here, we will limit ourselves to reproduce the main results.

\begin{figure}[h]
    \centering
    \includegraphics[width=\linewidth]{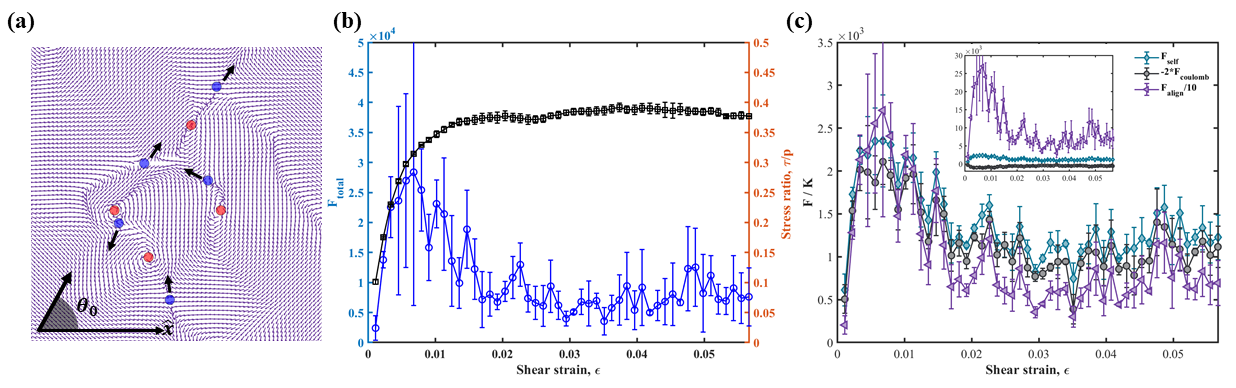}
    \caption{\textbf{(a)} Purple arrows are the nonaffine displacement field with red and blue disk respectively positive and negative TDs. Black arrows indicate the orientation of the negative TDs. \textbf{(b)} Comparison between the total energy as a function of $\epsilon$ and the stress ratio $\tau/p$. \textbf{(c)} Different contributions to the total defect energy as a function of $\epsilon$.The main panel shows the rescaled results, while the inset displays the raw values of the three energy components. }
    \label{fig:en}
\end{figure}
Let us consider two TDs with charges $w_1,w_2$, orientation angles $\theta_1,\theta_2$ and located at $(x_1,y_1)$ and $(x_2,y_2)$ respectively. We can define \cite{Selinger:619845} their relative orientation:
\begin{equation}
    \delta \theta= \theta_2-\theta_1 - \left[w_1 \tan^{-1} \left(\frac{y_2-y_1}{x_2-x_1}\right)-w_2 \tan^{-1} \left(\frac{y_2-y_1}{x_2-x_1}\right)\right].
\end{equation}
We notice that $\delta \theta \neq \theta_2-\theta_1$. This is because while the topological charges are addictive, the corresponding orientations have to be treated more carefully and cannot just be simply added. See \cite{Selinger:619845} for an extensive explanation of this fact.

The total free energy for a defect pair is generally written as $F_{12}=F_c+F_1+F_2+F_{\text{int}}+F_{\text{alignment}}$. Here, $F_c$ is the contribution of the regions inside the cores of the defects, that is in general unknown. $F_1$ and $F_2$ correspond to the free energy of each individual defect, given by:
\begin{equation}
    F_n= \pi K q_n^2 \log \left(\frac{L}{r_0}\right),\label{fn}
\end{equation}
where $K$ is the stiffness parameter associated with gradients in the orientational angle $\theta$ (elastic interactions), $L$ is the system size, $r_0$ the size of the defect core and $q_n$ the topological charge of the $n$-th defect.

The interaction term can be written as
\begin{equation}
    F_{\text{int}}(\mathbf{r}_1,\mathbf{r}_2)= - 2 \pi K q_1 q_2 \log \left(\frac{|\mathbf{r}_1-\mathbf{r}_2|}{L}\right)\label{int}
\end{equation}
and corresponds to a Coulomb-like interaction. As anticipated, this interaction is long-range and logarithmic, consistent with the behavior of electromagnetic interactions in two dimensions. The force on each defect is simply given by $f=-\partial F/\partial r_{12}$, where $r_{12}\equiv |\mathbf{r}_1-\mathbf{r}_2|$ is the distance between the defects. The resulting force is repulsive for charges of same sign and attractive otherwise. We notice that the attractive force between $w=+1$ and $w=-1$ defects explains the tendency of the defects to appear in pairs, near each other, as demonstrated by their structure in Fig. \ref{fig:2}(b). In fact, for a pair of defects with opposite sign the energy is minimized when they are close to each other.

Finally, the alignment term is given by
\begin{equation}
    F_{\text{alignment}}(\mathbf{r}_1,\mathbf{r}_2,\theta_1,\theta_2)=\frac{\pi K}{2}\,\delta\theta^2\frac{\log \left(|\mathbf{r}_1-\mathbf{r}_2|/(2 r_0)\right)}{\left[\log \left(|\mathbf{r}_1-\mathbf{r}_2|/r_0\right)\right]^2}.\label{ali}
\end{equation}
This term is more intriguing, as it describes an alignment interaction between topological defects (TDs) with an orientational mismatch $\delta \theta$. Here, $r_0$ denotes the defect core size. This term vanishes when the two defects are optimally aligned, i.e., when $\delta \theta = 0$. Moreover, it generates a torque on each defect, which can be directly estimated as $-\partial F_{\text{int}}/\partial \delta \theta$ \cite{Selinger:619845}. This term is rather similar to the interaction between two dislocations with misaligned Burgers vectors (see e.g. Eq. (17) in \cite{doi:https://doi.org/10.1002/9783527682300.ch2}).

Finally, within the pairwise interaction approximation, the total free energy of a given defect configuration can be obtained by summing the pairwise interaction energy $F_{12}$ over all distinct pairs of particles, i.e., for all $i \neq j$. Before proceeding, we note that the free energy discussed here does not include entropic contributions, which are crucial in many contexts, such as the well-known BKT transition and 2D melting. In granular systems, however, the athermal nature of the dynamics renders concepts like entropy and temperature ill-defined, making the inclusion of such terms nontrivial and deserving of separate consideration.

Additionally, we neglect the influence of external deformation, specifically, the energy transfer from shear strain to defect dynamics. A full treatment of these effects lies beyond the scope of this work. Nevertheless, as we will soon show, the defect-defect interactions alone already yield valuable physical insight.

Aware of these limitations, we proceed to compute the various contributions to the free energy. To do so, we assume that $r_0$ corresponds to the lattice grid size, as topological defects are defined only up to this microscopic scale. Additionally, due to the lack of precise knowledge of the stiffness parameter $K$, we set it to unity for simplicity. With these assumptions, we compute each free energy term at every strain step $\epsilon$ by tracking the positions and orientations of the topological defects. Positive defects, which lack a defined orientation, are excluded from the alignment term calculation.

To compute the free energy between defects, we first determine the phase angle $\theta_0$ of each defect~\cite{C7SM01195D}. The nonaffine displacement vector at the four lattice sites is used to identify the defect are used to calculate $\theta_0$. For a $+1$ defect, the direction of the nonaffine displacement at each vertex $\psi$ is computed with respect to the coordinate system defined by vectors pointing from the defect center to the corresponding lattice sites. The phase angle of the defect is then defined as $\theta_0 =\left< \psi \right>$, allowing us to distinguish between outward ($\theta_0 = 0$), inward ($\theta_0 = 0$), clockwise ($\theta_0 = 0$), and counter-clockwise ($\theta_0 = 0$) configurations.For $-1$ one, $\theta_0$ shows the orientation of the defects. We construct a symmetry tensor~\cite{C7SM01195D} as $T = \frac{\left< \nabla \vec{u} \right>}{\left| \left< \nabla \vec{u} \right>\right|}$ based on the two-fold symmetry of $-1$ defects. The eigenvector corresponding to the eigenvalue of $+1$ is selected as the orientation of the defect. The phase angle is then given by $\theta_0 = 2\psi$.

Then we need to confirm the constant in the equation of free energy. The system size is define by $L = 0.5*(x_{max}-x_{min}+y_{max}-y{min}$ and the core size of the defects is $r_0 = a$. Since all the displacement for computing the defects are obtained through interpolation from the discrete data, the core size is reasonable equal to the lattice size. When we choose different value of $r_0$ near $a$,  there is no unscalable difference in the result of free energy.  Given the current experimental resolution, the free energy within the defect core cannot be defined. Therefore, $F_n$ is temporarily excluded from the calculation.

According to the Eq.\ref{fn} Eq.\ref{int} and Eq.\ref{ali}, we can compute the free energy between each pair $F_{12}$ and total energy by summing $F_{12}$ for all of the pairs.

\subsection{Stability of the numerical protocol}
\begin{figure}[h]
    \centering
    \includegraphics[width=\linewidth]{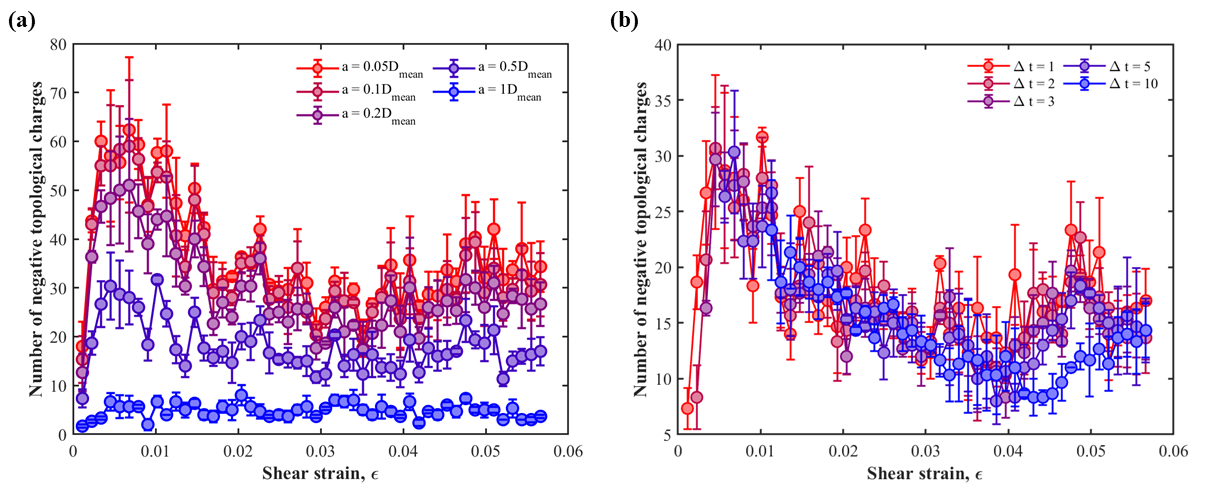}
    \caption{Number of topological defects with negative winding number (anti-vortices) as a function of the strain $\epsilon$. Panel \textbf{(a)} shows the results by varying the value of $a$ with $\Delta t=1$. Panel \textbf{(b)} show the same comparison by varying the time step used to compute the displacement field $\Delta t$ with $a=0.5D_{\text{mean}}$.}
    \label{si5fig}
\end{figure}

To compute the TD field, we interpolate the defects onto a square grid (see \textit{Methods} for details). To verify the reliability of our computation, we analyze how the number of defects varies with different grid spacings \( a \) in Fig.\ref{si5fig}(a). The trends of each curve are similar, exhibiting a linear region, a soft region, and a steady region as \( a \) increases from \( 0.05D \) to \( 0.5D \). When \( a = D \), the soft region becomes indistinct due to the low number of defects; with fewer than $10$ defects, the transition from round-like clusters to line-like clusters can no longer be observed clearly. However, the oscillatory behavior persists for all values of \( a \), which supports the validity of our analysis regarding the shear band formation mechanism.

In computing the non-radiative displacement field for TD identification, different time intervals can be chosen. The results in Fig.\ref{si5fig}(b) show that varying the time step has no effect on the number of defects.

\end{document}